\documentclass[12pt]{article}
\usepackage{amsmath}
\usepackage{graphicx}
\usepackage{enumerate}
\usepackage{natbib}
\usepackage{url} 

\usepackage{amsfonts,amsmath,amsthm,amssymb,graphicx,float,subcaption,  color,xcolor,natbib,url,hyperref,tikz,enumerate
}
\usepackage[shortlabels]{enumitem}
\usepackage{mathtools}

\graphicspath{{figures/}}
\allowdisplaybreaks

\definecolor{aqua}{RGB}{0,255,255}
\definecolor{fuchsia}{RGB}{255,0,255}

\newcommand{\bm}{\boldsymbol}

\newcommand{\bdelta}{\boldsymbol{\delta}}

\newcommand{\blind}{1}

\theoremstyle{plain}

\newtheorem{theorem}{Theorem}[section]
\newtheorem{lemma}[theorem]{Lemma}
\theoremstyle{remark}

\newtheorem{algorithm}{Algorithm}

\newcommand{\mb}{\mathbf}

\DeclareMathOperator*{\argmax}{arg\,max}

\def\mathbi#1{\textbf{\em #1}} 
\newcommand{\bI}{\boldsymbol{I}}

\newcommand{\by}{\boldsymbol{y}}

\begin{document}

	\def\spacingset#1{\renewcommand{\baselinestretch}%
		{#1}\small\normalsize} \spacingset{1}

\title{Information-based Optimal Subdata Selection for Clusterwise Linear Regression}
\if0\blind{\date{}}
\else{\author{ Yanxi Liu, John Stufken, and Min Yang \\
AbbVie Inc., George Mason University, and University of Illinois at Chicago}
  \date{\today}}
\fi
\maketitle

\begin{abstract}

Mixture-of-Experts  models are commonly used when there exists distinct clusters with different relationships between the independent and dependent variables.  Fitting such models for large datasets, however, is computationally virtually impossible. An attractive alternative is to use a subdata selected by ``maximizing" the Fisher information matrix. A major challenge is that no closed-form expression for the Fisher information matrix is available for such models. Focusing on clusterwise linear regression models, a subclass of MoE models, we develop a framework that overcomes this challenge. We prove that the proposed subdata selection approach is asymptotically optimal, i.e., no other method is statistically more efficient than the proposed one when the full data size is large.

\end{abstract}

\section{Introduction}

Modern information technologies, such as cloud computing, internet of things, social networking, etc., are drivers for exponential growth of the size of datasets. Size may now be measured by TB and even PB instead of MB and GB \cite{Cai2015TheEra}. While the extraordinary amount of data offers unprecedented opportunities for scientific discoveries and advancement, it also poses unprecedented challenges for analysis. These challenges are typically amplified by the complexity of the data and the speed with which it must be analyzed. A critical question for the statistics community is how to detect statistical relationships within high volumes of data with a complicated structure and turn it into actionable knowledge \cite{Buhlmann2016HandbookData}.

With large datasets, relationships between input and output variables may no longer be homogeneous. Linear models or generalized linear models, which are effective when relationships are homogeneous, may be inadequate in the era of big data. One strategy for dealing with heterogeneity is through Mixture-of-Experts (MoE) models. The rationale for MoE models is to uncover hidden clusters within the data, such that within each cluster relationships between input and output variables can be adequately modeled by a single regression or classification model. While any such regression or classification model may be inadequate for the entire dataset, it may be just fine for a more homogeneous cluster. Flexibility and interpretability of MoE models has resulted in their broad use in regression, classification, and fusion applications in healthcare, finance, surveillance, and recognition \cite{Yuksel2012TwentyExperts}.

The flexibility that MoE models provide goes however hand in hand with 
a high computational cost. The parameters of an MoE model are usually estimated using an EM algorithm, which requires a considerable computing time for each iteration when the data size is large. In addition, since the EM algorithm usually converges to a local rather than global optimum \cite{Balakrishnan2017StatisticalAnalysis, Wu1983OnAlgorithm}, different initial values of the parameters must be considered for better estimation results. This makes this approach inefficient and daunting for large datasets \citep{Makkuva2019BreakingAlgorithms}.

An attractive idea, which has received considerable attention for dealing with massive data ({\it full data}), is selection and analysis of a much smaller subset of the data ({\it subdata}). Using subdata of a much smaller size can overcome the computational burden, but reduces the information about the parameters contained in the original full data. For example.  \cite{Wang2019Information-BasedRegression,Cheng2020Information-basedRegression} proved that for linear and logistic regression models the information contained in the subdata selected by using popular random subsampling methods, including uniform random sampling, is asymptotically limited by the subdata size when the full data size becomes large. 

The Information-Based Optimal Subdata Selection (IBOSS) method \citep{Wang2019Information-BasedRegression}, which selects subdata judiciously, is computationally efficient and does not suffer from this limitation. For fitting a linear model, it is shown in \citep{Wang2019Information-BasedRegression} that, if each independent variable has a distribution in the domain of attraction of the generalized extreme value distribution, the variances of the estimators of the slope parameters based on analyzing subdata converge to zero when the full data size grows even though the subdata size is fixed. Studying properties for information-based subdata selection under generalized linear and nonlinear models is more challenging because there are no closed-form expressions for estimators and information matrices depend on the unknown parameters. \cite{Cheng2020Information-basedRegression} developed a two-stage IBOSS-based subdata selection algorithm for logistic regression models and proved, for selected cases, that the information matrices based on subdata of a fixed size increase with the full data size.

With the IBOSS strategy, the goal is to select subdata that maximizes a function of the Fisher information matrix for the parameters of interest. This is even more challenging for MoE models than for generalized linear and nonlinear models and requires novel ideas. 
The fact that there is no closed-form expression for the information matrix under an MoE model prevents the use of optimal design techniques for selecting efficient subdata, which is the strategy that was used for linear and logistic regression models. 


Focusing on the subclass of MoE models known as clusterwise linear regression models, we address this problem by using a surrogate matrix rather than the Fisher information matrix for guiding the subdata selection. We prove 
that the surrogate matrix is asymptotically equivalent to the information matrix under some mild conditions. We further prove that the statistical efficiency of the selection algorithm based on the surrogate matrix is asymptotically optimal,  i.e., there exists no other method with better statistical efficiency in terms of convergence rate when the full data size becomes large. 

In what follows, Section \ref{framework} introduces clusterwise linear regression models, while Section \ref{iboss_clr} presents the main results. Simulation studies and the analysis of real data are presented in Sections \ref{sim} and \ref{application}, respectively. Brief conclusions and possible future work are discussed in Section \ref{conclusion}. All technical details are presented in the Appendix.

\section{Mixture-of-Experts models and Clusterwise Linear Regression}
\label{framework}
Mixture-of-Experts models, which originated in the neural network literature \citep{Jacobs1991AdaptiveExperts}, are widely popular regression and classification models in machine learning due to their flexibility in modeling and appealing interpretation \citep{Masoudnia2014MixtureSurvey}. 
Rather than using a single model, MoE models are based on multiple models (or experts), which are mixed and combined, to provide great flexibility. MoE models assess how the data may be clustered into $G$ clusters so that separate regression or classification models can be used in each cluster. In combination with many current regression and classification algorithms, empirical evidence shows that MoE models are powerful tools to study relationships among variables in a variety of settings, including healthcare, finance, social science, etc. \cite{Yuksel2012TwentyExperts}. 

{Formally, let $(\mathbf{z}_i^T,y_i)$, $i=1,\ldots,N$, be independent, where $\mathbf{z}_i=(z_{i1},\ldots,z_{ip})^T$} is the covariate vector and $y_i$ is the response for the $i$th observation. We also use $\mathbf{x}_i=(1,\mathbf{z}_i^T)^T$. In a Mixture-of-Experts model, there are $G$ gate functions and $G$ regression models (experts). While $y_i$  is modeled by $\mathbf{x}_i$ through one of the experts, it is unknown which expert is employed.  A latent indicator vector can be used to describe the connection. Let $\bI_i=(I_{i1},\ldots,I_{iG})$, where
\begin{equation}\label{indicator}
I_{ig}=\begin{cases}
1 & \text{if the $g$th expert is employed,}\\
0 & \text{otherwise.}
\end{cases}.
\end{equation}
The likelihood of $I_{ig}=1$ is modeled by the $g$th gate function $P(I_{ig}=1|\mathbf{z}_i)$. While more complicated choices are possible, and sometimes advisable, a popular simple choice  is 
\begin{equation}\label{gate2}
\begin{split}
P(I_{ig}=1|\mathbf{z}_i)=\pi_g, \ g=1,\ldots,G,
\end{split}
\end{equation}
with $\sum_{g=1}^G\pi_g=1$. 

If $I_{ig}=1$, then we can model the response $y_i$ by $\mathbf{z}_i$ through the $g$th expert. The choice of the experts depends on the nature of the responses. For example, for a continuous response, a linear model may be appropriate for an expert; for a categorical response, experts may consist of generalized linear models. 

While MoE models were coined by \cite{Jacobs1991AdaptiveExperts}, the idea can be traced back to \cite{Fair1972MethodsDisequilibrium} and \cite{Hosmer1974MaximumLines}, where the experts are linear regression models. Such models, with the choice for the gate function as in (\ref{gate2}), were later called ``clusterwise linear regression'' (CLR) models \citep{Spath1979AlgorithmRegression} and have been widely applied in the social sciences, environmental studies, engineering, etc. \citep{Brusco2003MulticriterionValue, Bagirov2017PredictionRegressionapproach, Khadka2017ComprehensiveSystems}. 
Research on CLR models is still ongoing, especially on developing efficient algorithms for alleviating the computational burden \citep{DiMari2017ClusterwiseConstraints, Park2017AlgorithmsRegression}. 
If $(\mathbf{z}_i^T, y_i)$ belongs to the $g$th cluster, i.e., $I_{ig}=1$, then for a CLR model we write 
\begin{equation}\label{clr_def}
y_i =\mb{x}_i^T\bm{\beta}_g+\epsilon_i, \quad \epsilon_i \sim \mathcal{N}(0,\sigma^2_g),
\end{equation}
where $\bm{\beta}_g =(\beta_{0g},\beta_{1g},...,\beta_{pg})$ and for any two distinct $ g,g' \in \lbrace 1, ..., G\rbrace$, $\bm{\beta}_g \neq \bm{\beta}_{g'}$. In the remainder, we will focus on CLR models.

Analysis of a CLR model is primarily based on the maximum likelihood approach \citep{DeSarbo1988ARegression}. From \eqref{clr_def},  the distribution of $y_i$ is given by:
\begin{equation}
y_i \sim \sum_{g=1}^G\pi_g\phi(y_i|\mb{x}_i^T\bm{\beta}_g,\sigma^2_g)\qquad i=1,...,N
\end{equation}
where $\phi(\cdot|\mu,\sigma^2)$ is the density function for the normal distribution with mean $\mu$ and variance $\sigma^2$. For simplicity of notation, we will write $\phi_{ig}$ instead of $\phi(y_i|\mb{x}_i^T\bm{\beta}_g,\sigma^2_g)$. 
The loglikelihood function given $\by=(y_1,...,y_N)$ is then
\begin{equation}
\label{lik_clr}
l_{\by}=\sum\limits_{i=1}^N\log\left(\sum_{g=1}^G\pi_g\phi_{ig}\right).
\end{equation}

In contrast to a linear model, for a CLR model there is no closed-form expression for the MLE due to the summation over $g$ in the loglikelihood function (\ref{lik_clr}). In fact, without further restrictions there is an identifiability issue. Identifiability must be considered on equivalence classes of parameter vectors, so that two parameter vectors for which one can be obtained from the other by relabeling the clusters are considered to be equivalent. But even on such equivalence classes, identifiability is not automatic. For example, if the vectors $\mathbf{z}_i$ belong to a $(p-1)$-dimensional hyperplane, then the model is not even identifiable with $G=1$ (i.e., for a single expert).  
Fortunately, 
\cite{Hennig2000IdentifiabilityRegression} gave a sufficient condition for identifiability of CLR model (\ref{clr_def}). Let $\mathcal{Z}=\{\mathbf{z}_1,\ldots, \mathbf{z}_N\}$ and
\begin{equation}\label{min_cover}
h:=\min \left\{q: \mathcal{Z}\subset \bigcup_{i=1}^q H_i: H_i\in \mathcal{H}_{p-1} \right\},
\end{equation}
where $\mathcal{H}_{p-1}$ is the set of all hyperplanes of dimension $p-1$.
\begin{theorem}[Theorem 2.2, \cite{Hennig2000IdentifiabilityRegression}]\label{clr_idt}
	The CLR model in (\ref{clr_def}) is identifiable if $G<h$, where $G$ is the number of clusters and $h$ is defined in (\ref{min_cover}).	
\end{theorem}
The sufficient condition in Theorem \ref{clr_idt} is relatively mild. As long as the covariate set $\mathcal{Z}$ cannot be covered by the union of $G$ or fewer $(p-1)$-dimensional hyperplanes, identifiablity holds. 
Thus, loosely speaking, if the covariate values are sufficiently rich, then the sufficient condition holds and Model (\ref{clr_def}) is identifiable. For a big dataset, unless there are structural restrictions on the covariate values, we can expect identifiability to be satisfied.

For a CLR model, with the unobservable indicator vector, the EM algorithm is the workhorse for finding the MLE  \citep{Yuksel2012TwentyExperts}. For given initial values of the parameters, the MLE is obtained by alternating between the expectation and maximization steps until convergence. However, the EM algorithm typically converges to a local optimum, and not necessarily to the global optimum \citep{Wu1983OnAlgorithm,Balakrishnan2017StatisticalAnalysis}. We generally need to try a large number of initial values to improve its performance. In addition, $G$, the number of clusters, is unknown. We also need to try different values of $G$ to find the best one according to some criterion, such as AIC. Consequently, the computational cost for analyzing a CLR model is very high. For example, for simulated data of size $N=10^7$ and $p=10$ covariates, the computing time for fitting a linear regression model is around 0.2 seconds. In comparison, on the same platform, it takes around 470 seconds for fitting a CLR model with $G=5$ being known and only one initial value. The computation time can be significantly increased due to the inclusion of numerous initial parameter values, as well as the consideration of different values for $G$.
In this era, it is not uncommon for the data size to be in the millions or even billions, and the structure of the data can be more complicated. While high performance computing can be helpful, fitting MoE models for such big datasets still poses a tremendous challenge. This can be alleviated by using carefully selected subdata. 

As indicated in the Introduction, the IBOSS strategy for subdata selection has been proven, both theoretically and empirically, to select highly informative subdata. Extending this strategy to CLR models would be extremely appealing for big data analysis, and would drastically reduce computational costs by fitting a CLR model to subdata that retains as much information about the parameters as possible.

To describe the IBOSS strategy, let $\mathbi{I}(\mathbf{x}_i)$ denote the information matrix for the $i$th data point. With $\delta_i=1$ if the $i$th data point belongs to the subdata and $\delta_i=0$  otherwise, and under the assumption of independence, the information matrix based on the subdata is
\begin{equation}\label{iboss1}
\mathbi{I}(\bdelta)=\sum_{i=1}^N \delta_i \mathbi{I}(\mathbf{x}_i).
\end{equation}
We want to select $\bdelta=(\delta_1,\ldots,\delta_N)$, subject to
$\sum_{i=1}^N\delta_i=n$, to maximize, in some way, the information matrix in
(\ref{iboss1}). For this maximization we adopt the approach from optimal design of experiments \citep{Kiefer1959OptimumProblems}, where an interpretable function of $I(\bdelta)$ is used to induce a complete ordering of the information matrices.
If $\Psi$ is this function, then, subject to $\sum_{i=1}^N\delta_i=n$, we want to find subdata with indicator vector $\bdelta^{opt}$ so that
\begin{equation}\label{iboss2}
\bdelta^{opt}=\arg \max_{\bdelta}\Psi(I(\bdelta)).
\end{equation}
We will refer to any subdata selected in this way as IBOSS subdata. Algorithms for an approximate solution to this complex optimization problem can be based on the  characterization of an optimal design for the corresponding model. 

{For the CLR model, the information matrix for the $i$-th data point can be written as $\mathbi{I}(\mathbf{x}_i) = E(\frac{\partial l_{y_i}}{\partial \bm{\theta}}\frac{\partial l_{y_i}}{\partial\bm{\theta}^T})$, where
\begin{equation} \label{infclr}
l_{y_i}=log\big(\sum_{g=1}^G\pi_g\phi_{ig}\big),
\end{equation}
and $\phi_{ig}$ is defined in Equation (\ref{lik_clr}).} Here $\bm{\theta}$ is the vector of the $G(p+3)-1$ parameters, with $G(p+1)$ of them corresponding to the $\bm{\beta}_g$'s, $G$ to the $\sigma_g^2$'s, and $G-1$ to the $\pi_g$'s.
However, the summation structure within the $\log$ function in \eqref{infclr} prevents the derivation of a closed-form expression for $\mathbi{I}(\mathbf{x}_i)$. This in turn means that finding an optimal design is elusive, so that a new approach is needed for obtaining IBOSS subdata.

\section{Main results}
\label{iboss_clr}
\subsection{Bounding the Fisher information matrix}
\label{iboss_clr:info_clr}
Without a closed-form expression for $\mathbi{I}(\mathbf{x}_i)$, we define a matrix that is larger than the Fisher information matrix in terms of the Loewner order and that has a closed-form expression. We first expand a data point from ($\mathbf{z}_i^T, y_i$) to ($\mathbf{z}_i^T, y_i, \bI_i$), where $\bI_i = (I_{i1},...,I_{iG})^T$ and $I_{ig}$ is defined in (\ref{indicator}). (Despite using the notation $\bI$ both for an information matrix and a vector of latent indicators, the meaning will always be clear from the context.) 
The likelihood function under the CLR model for the complete $i$th data point ($\mathbf{z}_i^T, y_i, \bI_i$) is then given by
\begin{equation}
\label{complete_lik}
L_{C_i}=\prod_{g=1}^G\big[\phi_{ig}\pi_g\big]^{I_{ig}}.
\end{equation} 
Observe that $L_{C_i}=L_{y_i}\times L_{\mb{I}_i|y_i}$, where $L_{\mb{I}_i|y_i}$ is the likelihood function corresponding to the conditional distribution of $\mb{I}_i$  given $y_i$.

Corresponding to this factorization of the complete data likelihood function, we can write the Fisher information matrix for the $i$-th data point in the form of $\mathbi{I}(\mathbf{x}_i)=\mathbi{I}_{C_i}-\mathbi{I}_{M_i}$, where $\mathbi{I}_{C_i}$ is the complete data Fisher information matrix (or complete information matrix for short) based on the complete data likelihood function in \eqref{complete_lik} and {$\mathbi{I}_{M_i}$ is the information matrix corresponding to the  conditional distribution of $\mb{I}_i$ given $y_i$. The detailed derivation is presented in the Appendix. The expressions for $\mathbi{I}_{C_i}$ and $\mathbi{I}_{M_i}$ can be written as follows:
\begin{equation}
\label{complete_infor}
\mathbi{I}_{C_i} = 
\begin{pmatrix}
\mathbi{I}_{\bm{\beta}|C_i} & 0 & 0 \\
0 & \mathbi{I}_{\bm{\sigma^2}|C_i} & 0   \\
0 & 0 & \mathbi{I}_{\bm{\pi}|C_i}  
\end{pmatrix}
\end{equation}
where
\begin{equation}
    \mathbi{I}_{\bm{\beta}|C_i} =
        \begin{pmatrix}
           \pi_1\frac{\mathbf{x}_i\mathbf{x}_i^T}{\sigma_1^2} & &  &\mathbf{0}\\
             &\pi_2\frac{\mathbf{x}_i\mathbf{x}_i^T}{\sigma_2^2} & &\\
             &&\ddots& \\
            \mathbf{0}& & & & \pi_G\frac{\mathbf{x}_i\mathbf{x}_i^T}{\sigma_G^2}
        \end{pmatrix}\label{infor_beta_c},
\end{equation}
\begin{equation}
    \mathbi{I}_{\bm{\sigma}^2|C_i} =
        \begin{pmatrix}
           \frac{\pi_1}{2\sigma_1^4} & &  &\mathbf{0}\\
             &\frac{\pi_2}{2\sigma_2^4} & &\\
             &&\ddots& \\
            \mathbf{0}& & & & \frac{\pi_G}{2\sigma_G^4}
        \end{pmatrix}.
\end{equation} and
\begin{equation}
    \mathbi{I}_{\bm{\pi}|C_i} =
        \begin{pmatrix}
            \frac{1}{\pi_1}+\frac{1}{\pi_G} & \frac{1}{\pi_G} & ... &...& \frac{1}{\pi_G}\\
            \frac{1}{\pi_G} & \frac{1}{\pi_2}+\frac{1}{\pi_G} & \frac{1}{\pi_G} &...&\frac{1}{\pi_G}\\
            \vdots & & \ddots & & \vdots\\
            \frac{1}{\pi_G} & ... & ... & \frac{1}{\pi_{G-2}}+\frac{1}{\pi_G} & \frac{1}{\pi_G}\\
            \frac{1}{\pi_G} & ... & ... & \frac{1}{\pi_G} & \frac{1}{\pi_{G-1}}+\frac{1}{\pi_G}
        \end{pmatrix}.
\end{equation}
The expression for $\mathbi{I}_{M_i}$ is obtained by subtraction and 
its diagonal is given by
\begin{eqnarray*}
\left(diag(\mathbi{I}_{\bm{\beta}_1|M_i}),\ldots,diag(\mathbi{I}_{\bm{\beta}_G|M_i}),\mathbi{I}_{\bm{\sigma}_1^2|M_i},\ldots,\mathbi{I}_{\bm{\sigma}_G^2|M_i},\mathbi{I}_{\bm{\pi}_1|M_i},\ldots,\mathbi{I}_{\bm{\pi}_{G-1}|M_i}  \right),
\end{eqnarray*}
where, for a square matrix $\mathbi{A}=(a_{ij})$, the notation $diag(\mathbi{A})$ denotes the diagonal matrix with diagonal entries $a_{ii}$, 
\begin{eqnarray} \label{imi}
\begin{split}
    &\mathbi{I}_{\bm{\beta}_g|M_i}=\mathbb{E}\Big\lbrace w_{ig}(1-w_{ig})\frac{(y_i-\mb{x_i}^T\bm{\beta}_g)^2\mb{x_i}\mb{x_i}^T}{\sigma_g^4}\Big\rbrace,\\
    &\mathbi{I}_{\bm{\sigma}_g^2|M_i}=\mathbb{E}\Bigg\lbrace w_{ig}(1-w_{ig})\Bigg[-\frac{1}{2\sigma_g^2}+\frac{(y_i-\mb{x_i}^T\bm{\beta}_g)^2}{2\sigma_g^4}\Bigg]^2\Bigg\rbrace, \\
    &\mathbi{I}_{\bm{\pi}_g|M_i}=\mathbb{E}\Bigg\lbrace \frac{w_{ig}(1-w_{ig})}{\pi_g^2} + \frac{w_{iG}(1-w_{iG})}{\pi_G^2} +2 \frac{w_{ig}w_{iG}}{\pi_g\pi_G} \Bigg\rbrace,
\end{split}
\end{eqnarray} 
and $w_{ig}=\frac{\pi_g\phi_{ig}}{\sum_{l=1}^G\pi_l\phi_{il}}.$
A detailed derivation can be found in the Appendix.

\subsection{Basic Strategy}
\label{iboss_clr:basic}
Since we do not have a closed-form expression for $\mathbi{I}_{M_i}$, we face a significant hurdle in identifying subdata $\bdelta^*$ that maximizes $det(\mathbi{I}(\bdelta))$. To solve this dilemma, we first observe that for any $\bdelta$, in Loewner order,
\begin{equation} \label{eq:2}
\begin{split}
&\mathbi{I}(\bdelta)=\sum_{i\in \bdelta}\left(\mathbi{I}_{C_i}-\mathbi{I}_{M_i}\right) \leq \sum_{i\in \bdelta}\mathbi{I}_{C_i}, \text{ so that} \\
 &det(\mathbi{I}(\bdelta)) \leq  det(\sum_{i\in \bdelta}\mathbi{I}_{C_i}).
\end{split}
\end{equation}
The notation $\sum_{i\in \bdelta}$ simply means that we sum only over those $i$ for which $\delta_i=1$. {Based on \eqref{eq:2}, for a full data size $N$, if we have a strategy to find subdata $\bdelta^*_N$ such that (a) $\bdelta^*_N=\argmax_{\bdelta}det(\sum_{i\in \bdelta}\mathbi{I}_{C_i})$ and (b)  $\sum_{i\in \bdelta^*_N}\mathbi{I}_{C_i} - \mathbi{I}(\bdelta^*_N) \rightarrow 0$ when $N\rightarrow \infty$, then  the subdata $\bdelta^*_N$ is asymptotically optimal for maximizing $det(\mathbi{I}(\bdelta))$.}

{Thus, for a fixed $N$, we need to identify a subdata selection strategy that leads to a $\bdelta^*$ that gives an approximate solution for (a) and that satisfies the requirement in (b). 
Note that $det(\sum_{i\in \bdelta}\mathbi{I}_{C_i})$ is proportional to $\Big(det(\sum_{i\in \bdelta}{\mb{x}_i}\mb{x}_i^T)\Big)^G$, so that maximizing $det(\sum_{i\in \bdelta}\mathbi{I}_{C_i})$ is equivalently to maximizing $det(\sum_{i\in \bdelta}{\mb{x}_i}\mb{x}_i^T)$. 
\citet{Wang2019Information-BasedRegression} develop the computationally inexpensive IBOSS algorithm for obtaining an approximate solution to precisely this problem. 
}
\begin{algorithm}[Algorithm 1 \cite{Wang2019Information-BasedRegression}]
\label{algo_lr}
With $k$ as the subdata size and $p$ as the number of covariates, assume for simplicity that $r=k/(2p)$ is an integer. Execute the following steps:
\begin{enumerate}
  \item Select the data points with the $r$ smallest and $r$ largest values for the first covariate;
  \item Sequentially, for $j=2,...,p$, exclude the data points that were previously selected, and select the data points with the $r$ smallest and $r$ largest values for the $j$th covariate from the remaining data points.
\end{enumerate}
\end{algorithm}
Thus, $\bdelta^*$ obtained by using Algorithm~\ref{algo_lr} gives an approximate solution to the maximization of $det(\sum_{i\in \bdelta}\mathbi{I}_{C_i})$. We still need to show that it also satisfies 
$\sum_{i\in \bdelta^*}\mathbi{I}_{M_i}\rightarrow\mb{0}$ for $N \rightarrow \infty$. To circumvent that $\sum_{i\in \bdelta^*}\mathbi{I}_{M_i}$ does not have a closed-form expression, we will show that, in the Loewner ordering, it is dominated by a diagonal matrix that converges to $\mb{0}$ when $N \rightarrow \infty$. This would immediately imply that $\sum_{i\in \bdelta^*}\mathbi{I}_{M_i}$, which is a non-negative definite matrix, also converges to 0.

For $1 \leq g_1 , g_2 \leq G$, let 
\begin{eqnarray}
    \mb{f}_{i1}\big(g_1,g_2\big)=&diag\left(\mb{x}_i\mb{x}_i^T \int\Tilde{w}_i(g_1,g_2)\Delta^2_{\bm{\beta}_{ig_1}} dy_i\right), \label{eq:f1}\\
    f_{i2}\big(g_1,g_2\big)=&\int\Tilde{w}_i(g_1,g_2)\Delta^2_{\bm{\sigma}_{ig_1}} dy_i, \text { and}\label{eq:f2}\\
    f_{i3}\big(g_1,g_2\big)=&\int\Tilde{w}_i(g_1,g_2) dy_i, \label{eq:f3}
\end{eqnarray}
where $\Tilde{w}_i(g_1,g_2)=\sqrt{\pi_{g_1}\phi_{ig_1}\pi_{g_2}\phi_{ig_2}}$, $\Delta_{\bm{\beta}_{ig}}=\frac{y_i-\mb{x}_i^T\mb{\beta}_g}{\sigma_g^2}$ and $\Delta_{\bm{\sigma}_{ig}}=\frac{(y_i-\mb{x}_i^T\mb{\beta}_g)^2-\sigma_g^2}{2\sigma_g^4}$. 
We consider  
\begin{eqnarray}
    \mathbi{Q}^i=diag \left( blkdiag\left(
\mathbi{Q}^i_{\bm{\beta}}, \mathbi{Q}^i_{\bm{\sigma^2}}, \mathbi{Q}^i_{\bm{\pi}} \right) \right), 
\end{eqnarray}
where, for matrices or scalars $\mathbi{A}_\ell$, $\ell = 1,...,L$, which can be of different dimensions, $blkdiag(\mathbi{A}_1,...,\mathbi{A}_L)$ denotes the block diagonal matrix with $\mathbi{A}_1, ..., \mathbi{A}_L$ along the diagonal, 
\begin{eqnarray}
    \mathbi{Q}^i_{\bm{\beta}}=
blkdiag\left(\mathbi{Q}^{i}_{\bm{\beta}_1},\ldots,\mathbi{Q}^{i}_{\bm{\beta}_G}\right)
 \label{eq:4}
\end{eqnarray} 
with  $\mathbi{Q}^{i}_{\bm{\beta}_g}=\frac{1}{2}\sum\limits_{g^*:g^* \neq g}\mb{f}_{i1}\big(g,g^*\big)$,
\begin{eqnarray}
    \mathbi{Q}^i_{\bm{\sigma^2}}=
blkdiag \left({Q}^{i}_{{\sigma}_1^2},\ldots, {Q}^{i}_{{\sigma}_G^2} \right),
 \label{eq:5}
\end{eqnarray}
 with ${Q}^{i}_{{\sigma}_g^2}=\frac{1}{2}\sum\limits_{g^*:g^* \neq g}f_{i2}\big(g,g^*\big)$, and
\begin{eqnarray}
    \mathbi{Q}^i_{\bm{\pi}}=
blkdiag\left({Q}^{i}_{{\pi}_1},\ldots, {Q}^{i}_{{\pi}_{G-1}}\right),
\label{eq:6}
\end{eqnarray}
with, for $1 \le g \le G-1$,  ${Q}^{i}_{{\pi}_g}=\frac{1}{2} \Bigg(\sum\limits_{g^*:g^*\neq g} \frac{f_{i3}(g,g^*)}{\pi_g^2}\Bigg) + \frac{1}{2} \Bigg(\sum\limits_{g^*:g^*\neq G}\frac{f_{i3}(G,g^*)}{\pi_G^2}\Bigg) + \frac{f_{i3}(g,G)}{\pi_g\pi_G}$.
With this notation, the following theorem holds. 
\begin{theorem}
\label{surrogate}
Assuming that $y_i \sim \sum_{g=1}^G\pi_g\phi(\mb{x}^T_i\mb{\beta}_g,\sigma_g^2)$, then, for any $\bdelta$, it holds that $diag(\sum_{i\in \bdelta}\mathbi{I}_{M_i})\leq \sum_{i\in \bdelta} \mathbi{Q}^i$ in terms of the Loewner ordering.
\end{theorem}
With the help of Theorem \ref{surrogate}, we can show that  $\sum_{i\in \bdelta}\mathbi{I}_{M_i}$ vanishes under certain conditions for subdata selected by Algorithm~\ref{algo_lr}.

\subsection{Main Theorems}
\label{iboss_clr:main}

Let $\bm{\mu}_z=(\mu_{z1},...,\mu_{zp})^T$ and $\bm{\Sigma}_z=\bm{\Phi}_z\bm{\rho}\bm{\Phi}_z$ be a full rank covariance matrix, where $\bm{\Phi}_z=blkdiag(\sigma_{z1},\ldots,\sigma_{zp})$ is a diagonal matrix  of standard deviations and $\bm{\rho} = (\rho_{jj'})_{p \times p}$ is a correlation matrix. 

\begin{theorem}
\label{main1}
Let $\mathbf{z}_1, ..., \mathbf{z}_N$ be iid, where $\mathbf{z}_i=(z_{i1},z_{i2},...,z_{ip})^T$. Assuming that $y_i \sim \sum_{g=1}^G\pi_g\phi(\mb{x}^T_i\bm{\beta}_g,\sigma_g^2)$, where $\mb{x}^T_i=(1,\mb{z}_i^T)^T$, and $\bdelta^*$ corresponds to subdata selected by Algorithm~\ref{algo_lr}, then 
$\sum_{i\in \bdelta^*}\mathbi{I}_{M_i} \xrightarrow{\mathbb{P}} \bm{0}_{(Gp+3G-1)\times (Gp+3G-1)} $ when $N \rightarrow \infty$ under one of the following conditions:
\begin{item}
\item[(a)] $\mathbf{z}_i \sim \mathbi{N}(\bm{\mu}_z,\bm{\Sigma}_z)$ and for any triplet $(g , g', j)$ with $g,g' \in \{1,...,G\}, g \not = g'$ and $j \in \{1,...,p\}$, it holds that $\sum\limits_{l=1}^p\rho_{lj}\sigma_{zj}(\beta_{g,l}-\beta_{g',l}) \neq 0 $;
\item[(b)] $\mathbf{z}_i \sim \mathbi{LN}(\bm{\mu}_z,\bm{\Sigma}_z)$ and for any triplet $(g , g' , j)$ with $g,g' \in \{1,...,G\}, g \not = g'$ and $j \in \{1,...,p\}$, it holds that  $ \beta_{g,j}-\beta_{g',j} \neq 0$ and $ \sum\limits_{l \in \mathcal{L}_{\min,j}}\big(\beta_{g,l}-\beta_{g',l}\big) \neq 0$, where $\mathcal{L}_{min,j}=\big\lbrace l \ \big| \  \rho_{lj}=\rho_{\min,j} \ ; \ l=1,...,p \big\rbrace$ and 
$\rho_{\min,j} = \min\limits_l \rho_{lj} < 0$.
\end{item}
\end{theorem}
The condition in (a) on the parameter space $\Theta \subset \mathbb{R}^{G(p+3)-1}$ is rather mild. If the condition is not satisfied, the parameter space will be reduced to a lower-dimensional subspace. The condition in (b) is more restrictive due to the requirement $\rho_{\min,j}<0$, which is needed for technical reasons. 

In view of Theorem~\ref{main1}, and guided by the basic strategy formulated at the beginning of this subsection, we propose the following algorithm for fitting a CLR model for a large dataset:
\begin{algorithm} 
With $k$ as the subdata size and $p$ as the number of covariates, assume for simplicity that $r=k/(2p)$ is an integer. Execute the following steps:
\begin{enumerate}
  \item Run Algorithm \ref{algo_lr} to select the subdata $\bdelta^*$;
  \item Using the EM algorithm, fit the CLR model using the subdata selected in Step 1. 
\end{enumerate}
\label{alg_clr}
\end{algorithm}

While Theorem~\ref{main1} establishes that the basic strategy works, it sheds no light on the statistical or computational efficiency of Algorithm~\ref{alg_clr}. The next theorem and the empirical results in Sections \ref{sim} and \ref{application} show that the statistical efficiency of Algorithm~\ref{alg_clr} is asymptotically optimal. We will return to the computational efficiency in Section~\ref{sim}.

\begin{theorem}
Let $\mathbf{z}_1, ..., \mathbf{z}_N$, where $\mathbf{z}_i=(z_{i1},z_{i2},...,z_{ip})$, be iid and let $k$ be the size of the subdata. Assume that $r=k/(2p)$ is an integer. 
Let  $y_i \sim \sum_{g=1}^G\pi_g\phi(\mb{x}^T_i\bm{\beta}_g,\sigma_g^2)$, where $\mb{x}^T_i=(1,\mb{z}_i^T)^T$, and let $\hat{\bm{\beta}}_g^{\bdelta^*}$ be the estimator of $\bm{\beta}_g$,  $g=1,\ldots,G$, under Algorithm~\ref{alg_clr}. 
\begin{item}
\item[(a)] If $\mathbf{z}_i \sim \mathbi{N}(\bm{\mu}_z,\bm{\Sigma}_z)$ and $\sum\limits_{l=1}^p\rho_{lj}\sigma_{zj}(\beta_{g,l}-\beta_{g',l}) \neq 0 $ for any triplet $(g , g', j)$ with $g,g' \in \{1,...,G\}, g \not = g'$ and $j \in \{1,...,p\}$, then, when $N \rightarrow \infty$,
\begin{eqnarray}
\label{asymp_1}
V(\bm{A}_N\hat{\bm{\beta}}^{\bdelta^*}_{g}) \rightarrow \frac{\sigma^2_g}{\pi_g}\begin{pmatrix}
 \frac{1}{k} &  \bm{0}\\
 \bm{0} & \frac{1}{4r}(\bm{\Phi}_z\bm{\rho}^2\bm{\Phi}_z)^{-1}
\end{pmatrix} 
\end{eqnarray}
where $\bm{A}_N=blkdiag(1,\sqrt{\log N}, \ldots, \sqrt{\log N})$.
\item[(b)] If $\mathbf{z}_i \sim \mathbi{LN}(\bm{\mu}_z,\bm{\Sigma}_z)$ and for any triplet $(g , g' , j)$ with $g,g' \in \{1,...,G\}, g \not = g'$ and $j \in \{1,...,p\}$, it holds that $ \beta_{g,j}-\beta_{g',j} \neq 0$ and $ \sum\limits_{l \in \mathcal{L}_{\min,j}}\big(\beta_{g,l}-\beta_{g',l}\big) \neq 0$, where $\mathcal{L}_{min,j}=\big\lbrace l \ \big| \  \rho_{lj}=\rho_{\min,j} \ ; \ l=1,...,p \big\rbrace$ and 
$\rho_{\min,j} = \min\limits_l \rho_{lj} < 0$, then, when $N \rightarrow \infty$,
\begin{eqnarray}
V(\bm{B}_N\hat{\bm{\beta}}^{\bdelta^*}_{g}) \rightarrow \frac{2\sigma^2_g}{k\pi_g}\begin{pmatrix}
 1 & & -\bm{\nu}^T\\
-\bm{\nu} & & p\bm{\Psi}+\bm{\nu}\bm{\nu}^T 
\end{pmatrix}
\label{asymp_2},
\end{eqnarray}
where $\bm{B}_N=blkdiag\Big(1,\exp(\sigma_{z1}\sqrt{2\log N}),...,\exp(\sigma_{zp}\sqrt{2\log N})\Big)$, $\bm{\nu}=\Big(e^{-\mu_{z1}},...,e^{-\mu_{zp}}\Big)^T$,\\ and $\bm{\Psi}=blkdiag\Big( e^{-2\mu_{z1}},...,e^{-2\mu_{zp}} \Big)$.
\end{item}
\\In addition, in both cases, the convergence rate for $\hat{V}(\hat{\bm{\beta}}_{g,j}^{\bdelta^*})$, $g=1,\dots,G$, is asymptotically optimal.
\label{asymp_thm}
\end{theorem}


\textbf{Remark:}  Theorem \ref{asymp_thm} delivers two important messages. First, in terms of statistical efficiency, the convergence rate of the proposed algorithm is asymptotically optimal. Second, it shows that for a fixed subdata size, we retain rich information about the regression parameters in the subdata. These desirable theoretical properties are confirmed by simulation studies in Section \ref{sim}.

Notice that, while the $\rho_{\min,j}<0$ condition in (b) is more restrictive due to the technical reasons, the simulation studies in Section \ref{sim} indicate the asymptotic results still hold even this condition is not satisfied.

\section{Simulation Studies}
\label{sim}
This section presents simulation studies to evaluate the performance of the proposed algorithm in terms of mean squared error for parameter estimation and computing time. We compare our method to obtaining subdata by random sampling (Random) to analyzing the full data (Full), with the latter serving as a benchmark.

In this simulation, we assume that the number of clusters $G$ is known. The full data of size $N$ is generated from a CLR model with $p=10$, $G=5$, and $\pi_1=0.1$, $\pi_2=0.1$, $\pi_3=0.2$, $\pi_4=0.3$, and $\pi_5=0.3$. We set $\sigma_g=g$ and $\bm{\beta}^T_g=\Big(\beta_{g,0},\bm{\beta}_{g,1}^T\Big)$ where $\bm{\beta}_{g,1}^T=\bigg(g,\, g+1,\ldots, g+9\bigg)$ and $\beta_{g,0}=g$  for $g=1,2,3,4,5$. For the covariance matrix of the covariates, $\bm{\Sigma}_z$, we use $\bm{\Sigma}_{z_{ij}}=0.5^{\bm{1}_{\lbrace i \neq j\rbrace}}$. The covariate vectors $\bm{z}_i$ are independent and identically distributed as $N(\bm{0},\bm{\Sigma}_z)$ or $LN(\bm{0},\bm{\Sigma}_z)$. For each of these, the simulation is repeated 100 times and empirical mean squared errors (MSE) for estimating the intercept and slope parameters are computed as $MSE_{\beta_{0}}=\frac{1}{100}\sum\limits_{s=1}^{100}\sum\limits_{g=1}^5(\hat{\beta}^{(s)}_{g,0} - \beta_{g,0})^2$ and $MSE_{\bm{\beta}_{1}}=\frac{1}{100}\sum\limits_{s=1}^{100}\sum\limits_{g=1}^5\Big|\Big|\hat{\bm{\beta}}^{(s)}_{g,1} - \bm{\beta}_{g,1} \Big|\Big|_2^2$, respectively. 


 \begin{figure}\centering
 \subfloat[MSE]{\includegraphics[width=.5\linewidth]{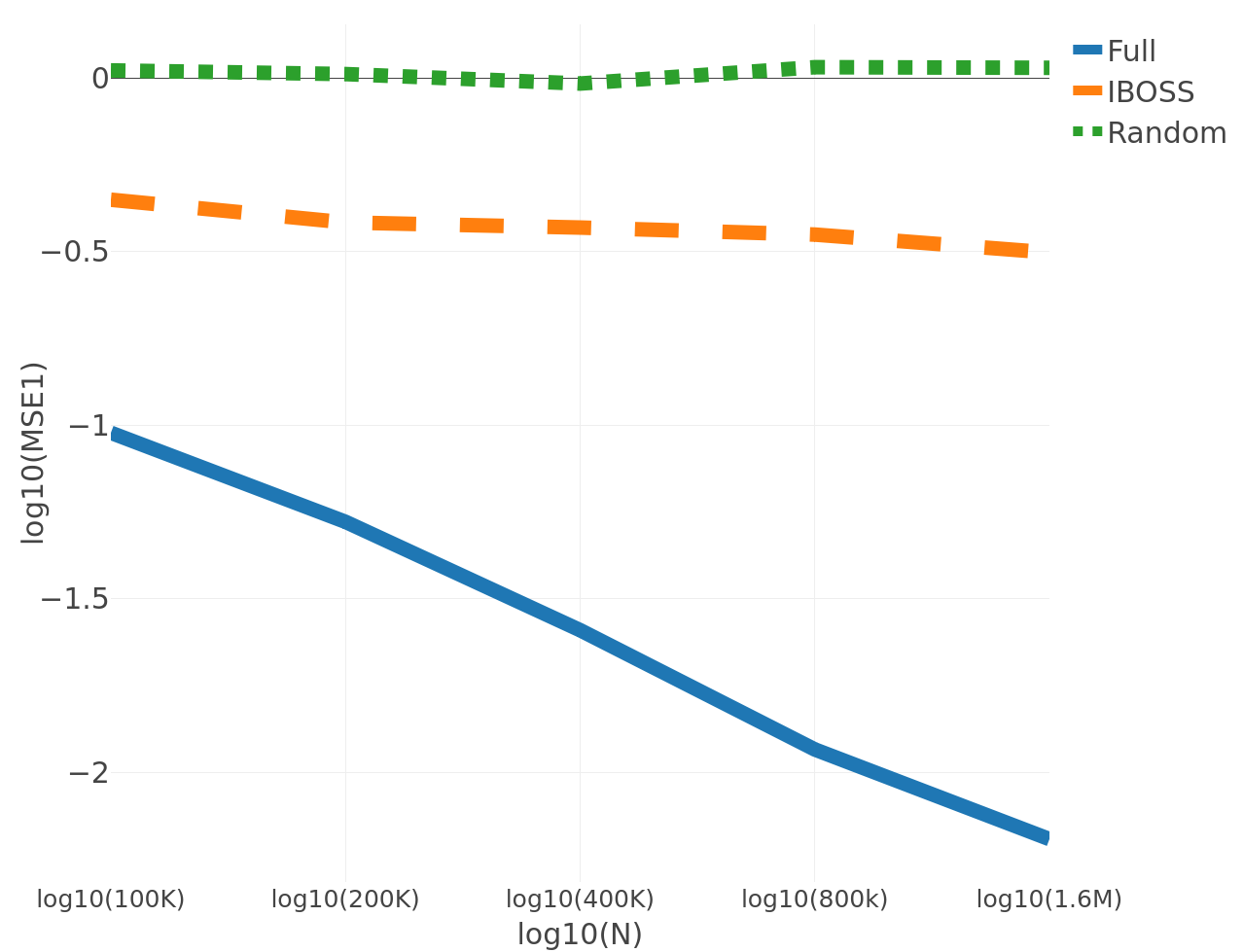}}\hfill
 \subfloat[CPU Time (in seconds)]{\includegraphics[width=.5\linewidth]{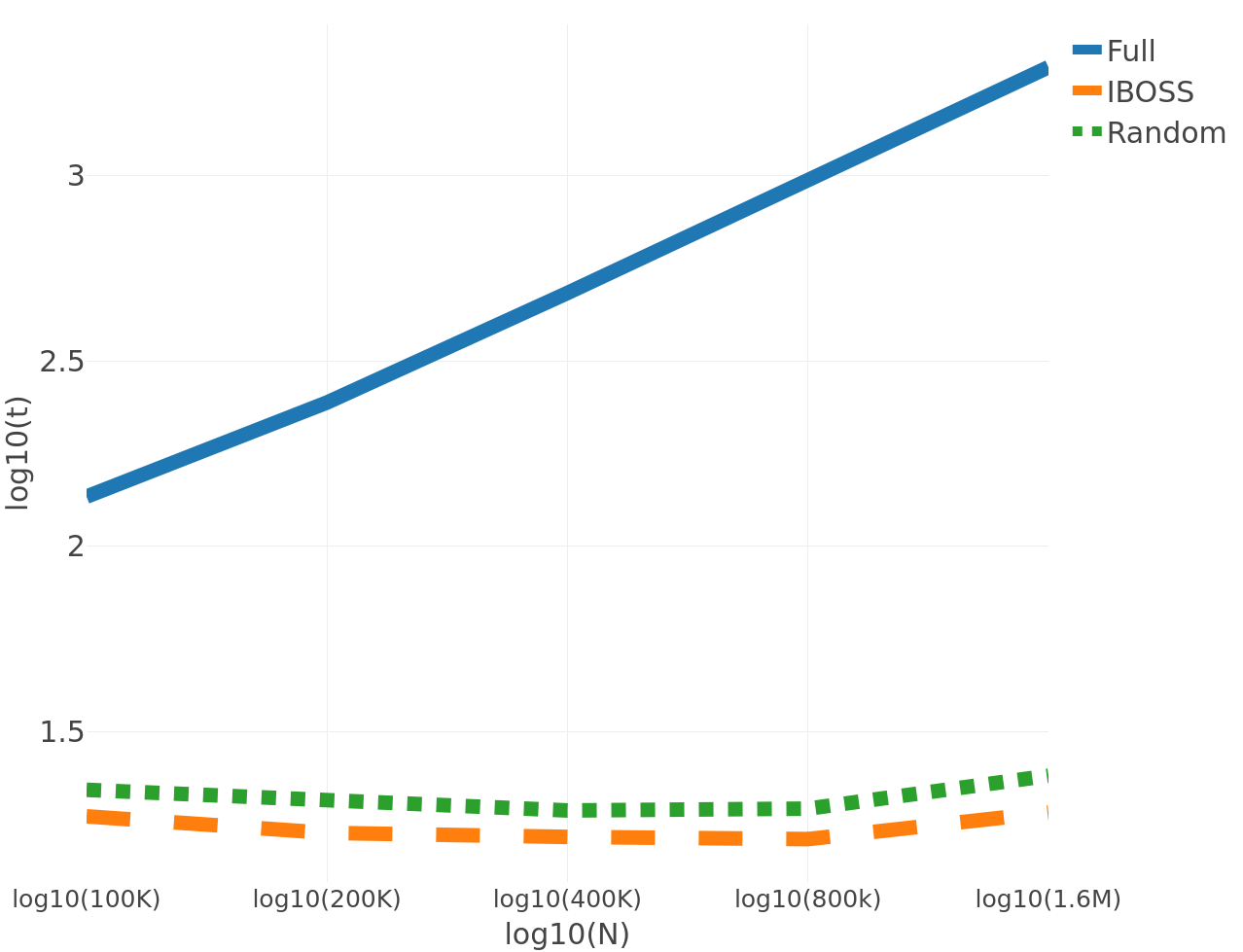}}
 \caption{Comparing different methods for estimating slope parameters when covariates are multivariate normal, subdata size $k=10000$, and full data size $N$ varies}
 \label{normal}
 \end{figure}

 \begin{figure}\centering
 \subfloat[MSE]{\includegraphics[width=.5\linewidth]{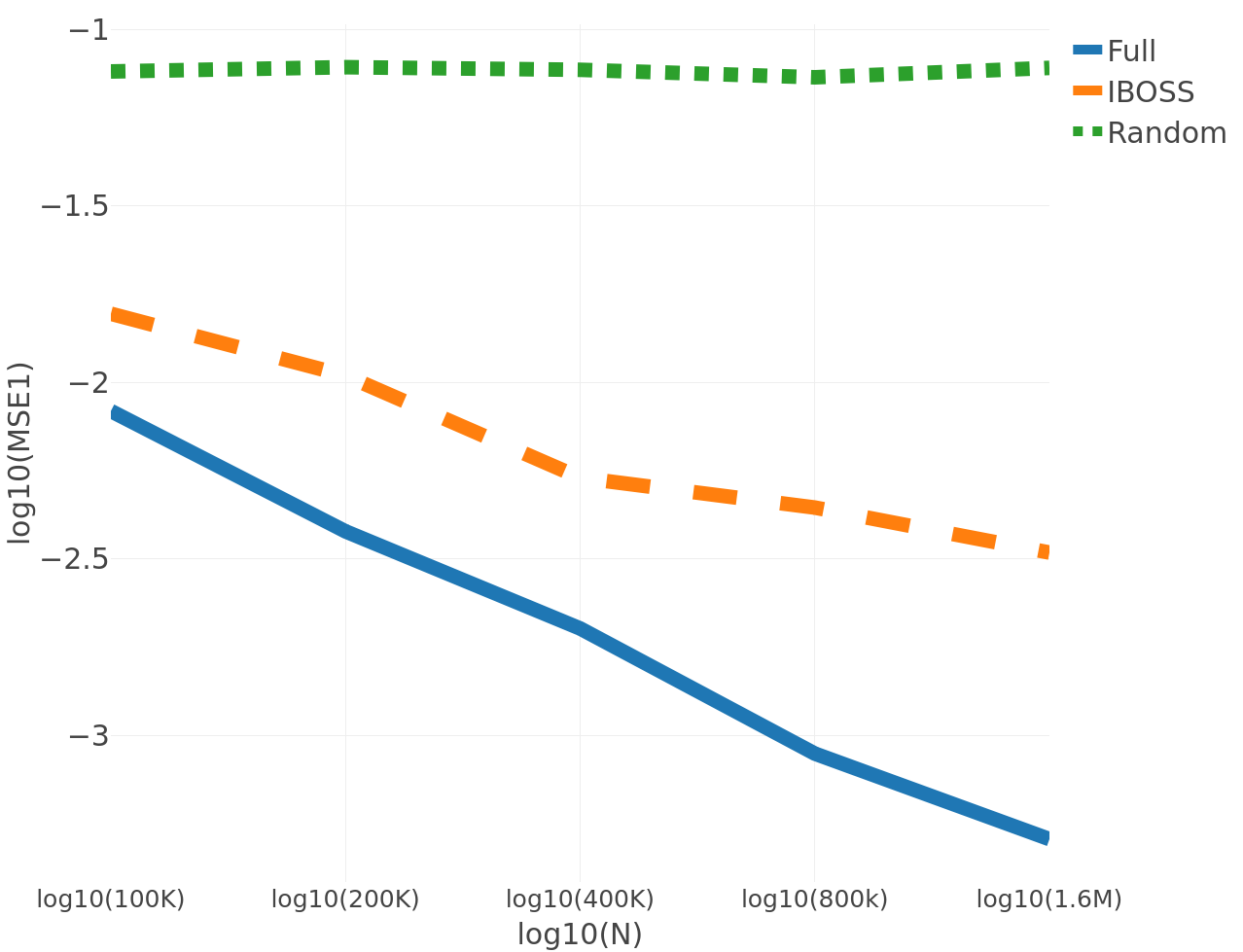}}\hfill
 \subfloat[CPU Time (in seconds)]{\includegraphics[width=.5\linewidth]{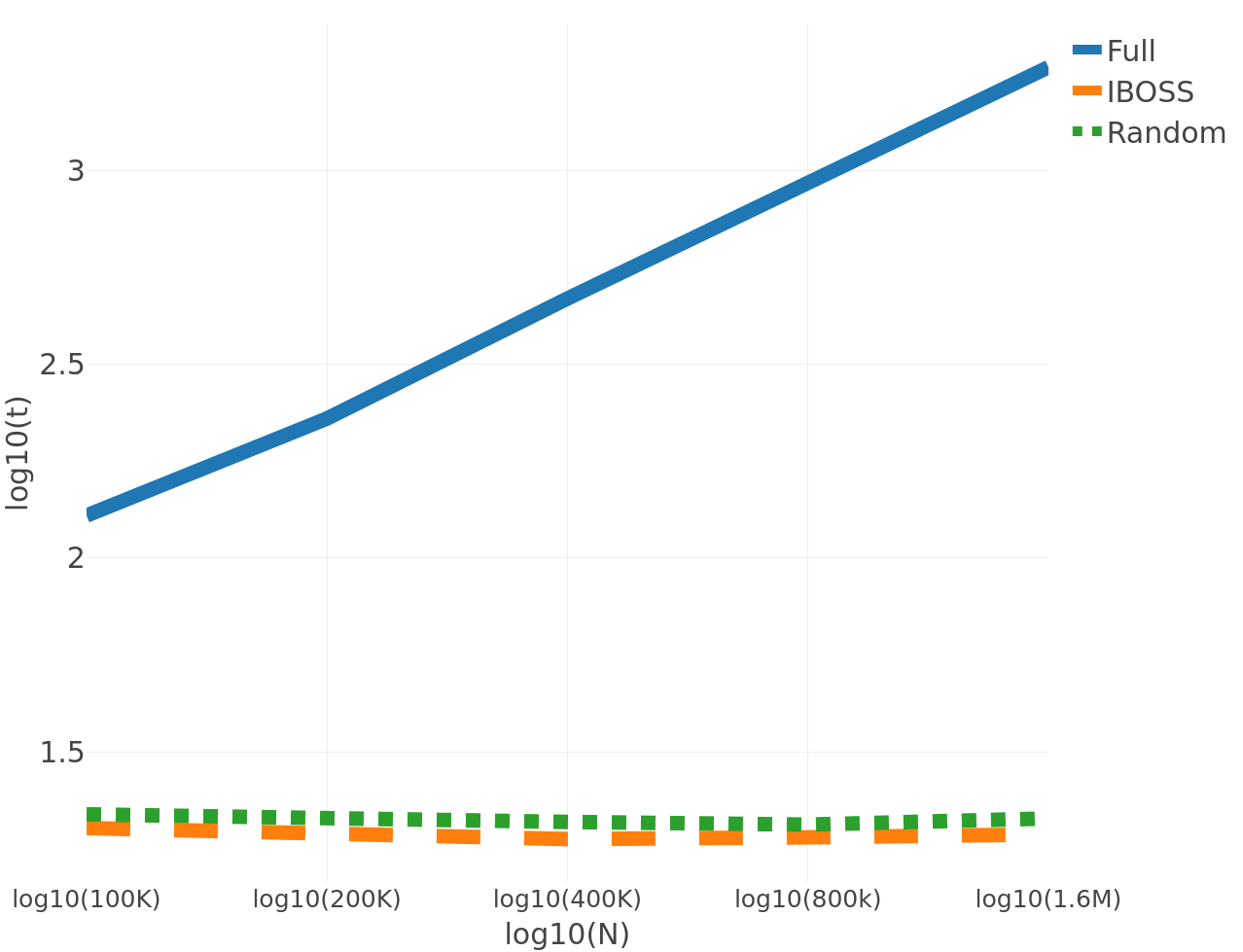}}
 \caption{Comparing different methods for estimating slope parameters when covariates are multivariate lognormal, subdata size $k=10000$, and full data size $N$ varies}
 \label{lognormal}
 \end{figure}



\begin{figure}\centering
 \subfloat[Normal]{\includegraphics[width=.5\linewidth]{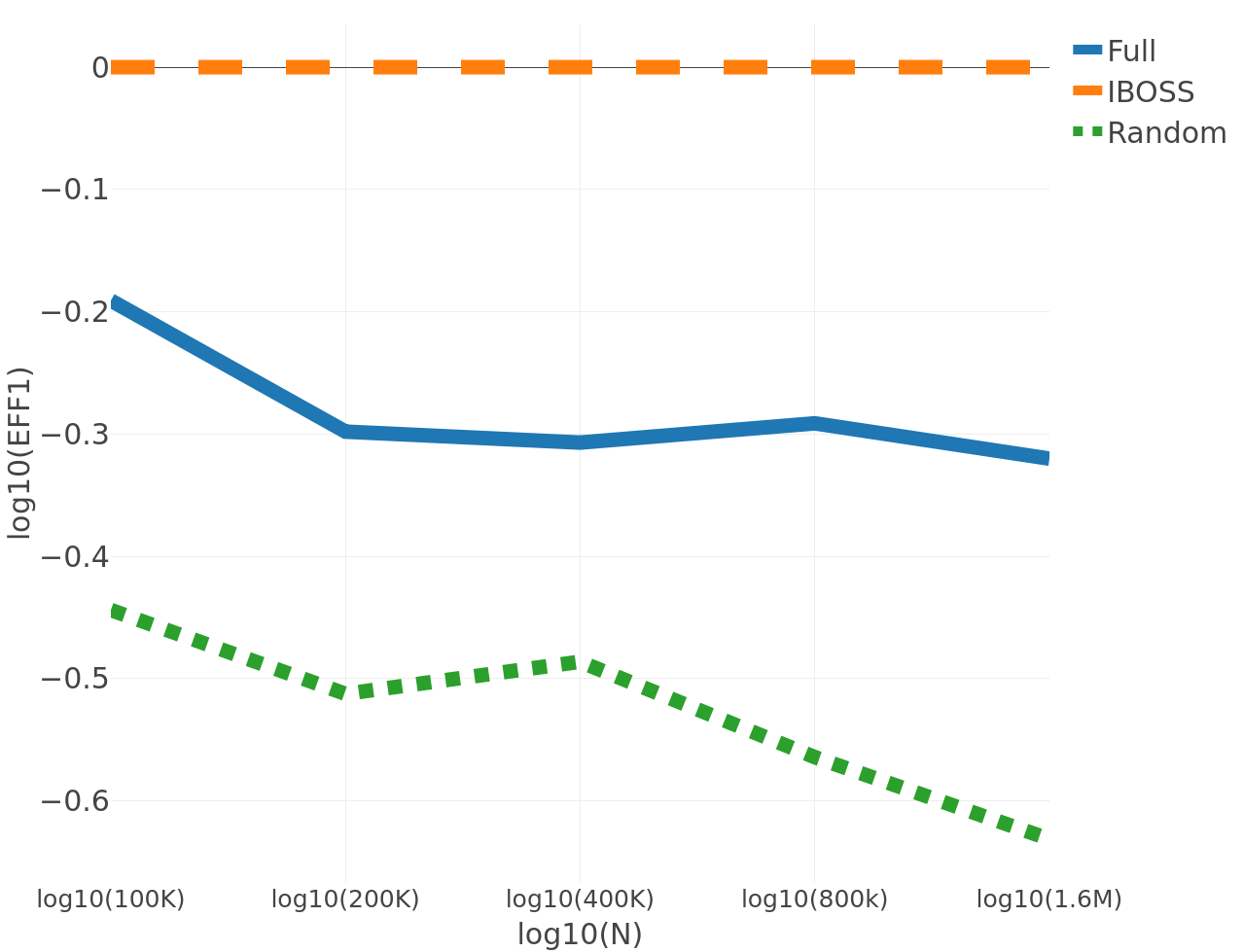}}\hfill
 \subfloat[LogNormal]{\includegraphics[width=.5\linewidth]{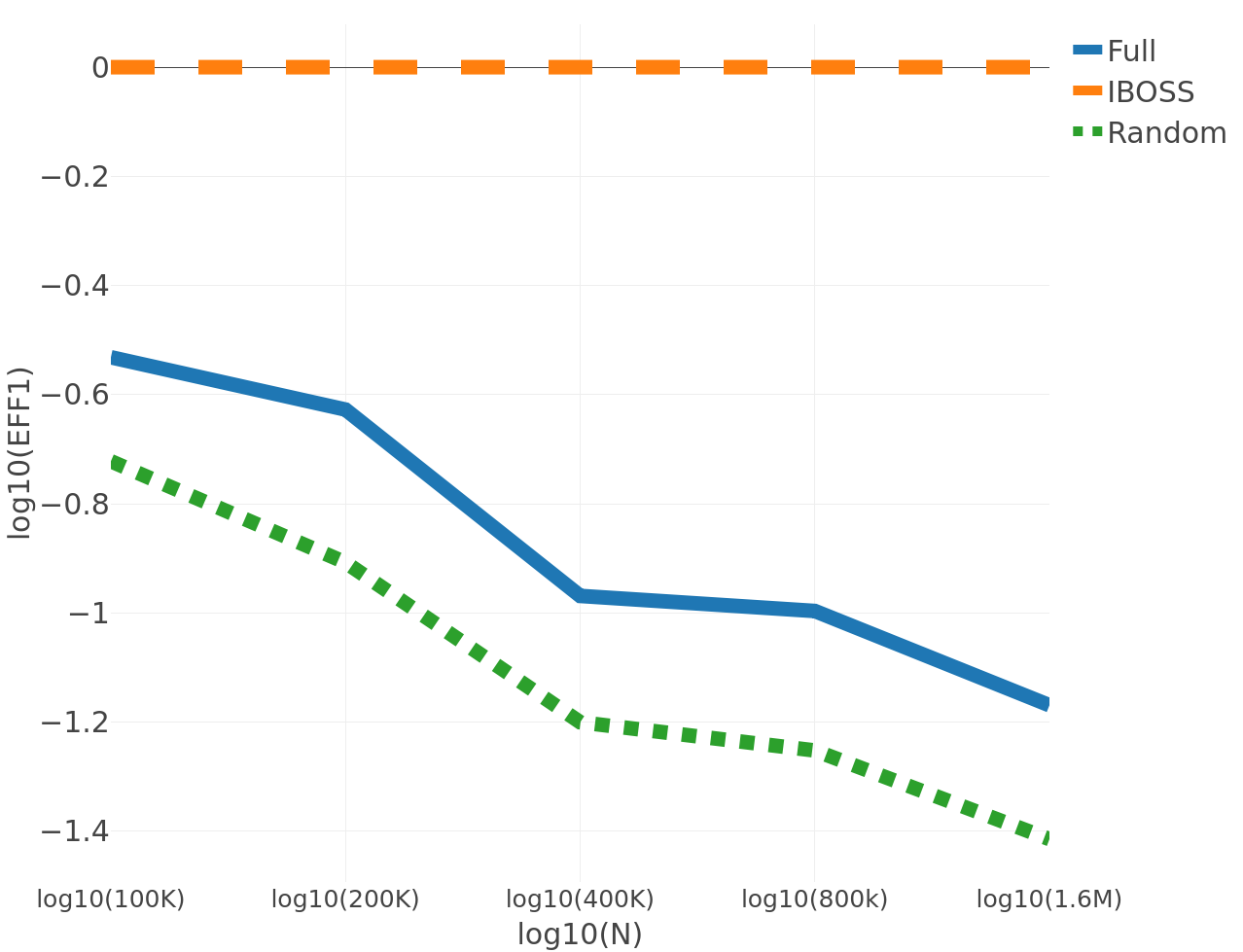}}
 \caption{Relative Efficiencies of different methods for slope parameters, subdata size $k=10000$, and full data size $N$ varies}
 \label{normalandlognormal}
 \end{figure}

For full data sizes $N=  10^5, 2\times 10^5,  4\times 10^5,  8\times 10^5,  1.6\times 10^6$ with fixed subdata size $k=10000$, Figures~\ref{normal} and ~\ref{lognormal} display the comparison of different methods for estimating the slope parameters with multivariate normal and lognormal covariate distributions, respectively. In both Figure~\ref{normal} (a) and Figure~\ref{lognormal} (a), it is seen that the MSE for the IBOSS method decreases as the full data size increases. This is consistent with the result of Theorem~\ref{asymp_thm}.



Both Figure~\ref{normal} (b) and Figure~\ref{lognormal} (b) show the computing time $t$ (in seconds) for each method across different full data sizes. Computing times were obtained by running Julia 1.8.5 code on an Inspiron 16 plus with 32GB ram and Intel Core i7-12700H. The computing times for FULL increase linearly with the full data sizes on the log-scale. The computing time (including subdata selection and data analysis) for the IBOSS and Random methods are virtually constant across different full data sizes. The computing time for IBOSS is even shorter than that for Random, which is due to faster convergence of the EM algorithm with IBOSS subdata than with Random subdata. 

To address the trade-off between computing time and statistical efficiency, one could define the relative efficiency for method $A$ compared to IBOSS as 
$$Eff_A=\frac{MSE_{IBOSS}/MSE_{A}} { Time_{A}/Time_{IBOSS}},$$
where $Time_{A}$ is the CPU time for method $A$. If $Eff_A=0.5$, say, one could think of this as IBOSS only needing half the CPU time of method $A$ to achieve the same MSE, or as IBOSS achieving half the MSE of method $A$ with the same CPU time. Figure~\ref{normalandlognormal} presents these relative efficiencies (on a log-scale) for Random and Full for different full data sizes $N$ and subdata size $k=10000$. Figure~\ref{normalandlognormal} shows that the relative efficiencies for Random and Full are smaller if covariates follow the multivariate Lognormal distribution. Also, over the range studied here, the relative efficiencies for Random and Full tend to decrease when the full data size $N$ increases.

\section{Application on Structural Protein Data}
\label{application}
In this section, we compare the performance of different methods on Structural Protein Data that was originally made available through the PBD.\footnote{Data is retrieved from https://www.kaggle.com/shahir/protein-data-set} Biomedical researchers can use the PDB to investigate various illnesses and develop new medicines and solutions that are vital to human existence. In this data set, we analyze the relationships between two variables: the explanatory variable, Structure Molecular Weight, and the response variable, Residue Count. After data cleaning, the full data size is $N=140,913$. 


{{Considering the choice $G=3$, the estimated parameters for two of the three clusters exhibit remarkable similarity. This observation strongly suggests that $G=2$ is a more suitable choice.}}  To compare {this method} to Random, we compute the MSEs for the slope parameters by using 500 bootstrap samples of size $n$, using $n=2\times 10^4, 4\times10^4,$ and $8\times10^4$. Subdata of size $k=1000$ is used, both for IBOSS and Random. 
The MSEs for the slopes are defined as in Section~\ref{sim} except that we replace $\bm{\beta}_{g,1}$ by the slope estimates from the full data, $\hat{\bm{\beta}}_{g,1}^{FULL}$.

 \begin{figure}\centering
 \subfloat[MSE]{\includegraphics[width=.5\linewidth]{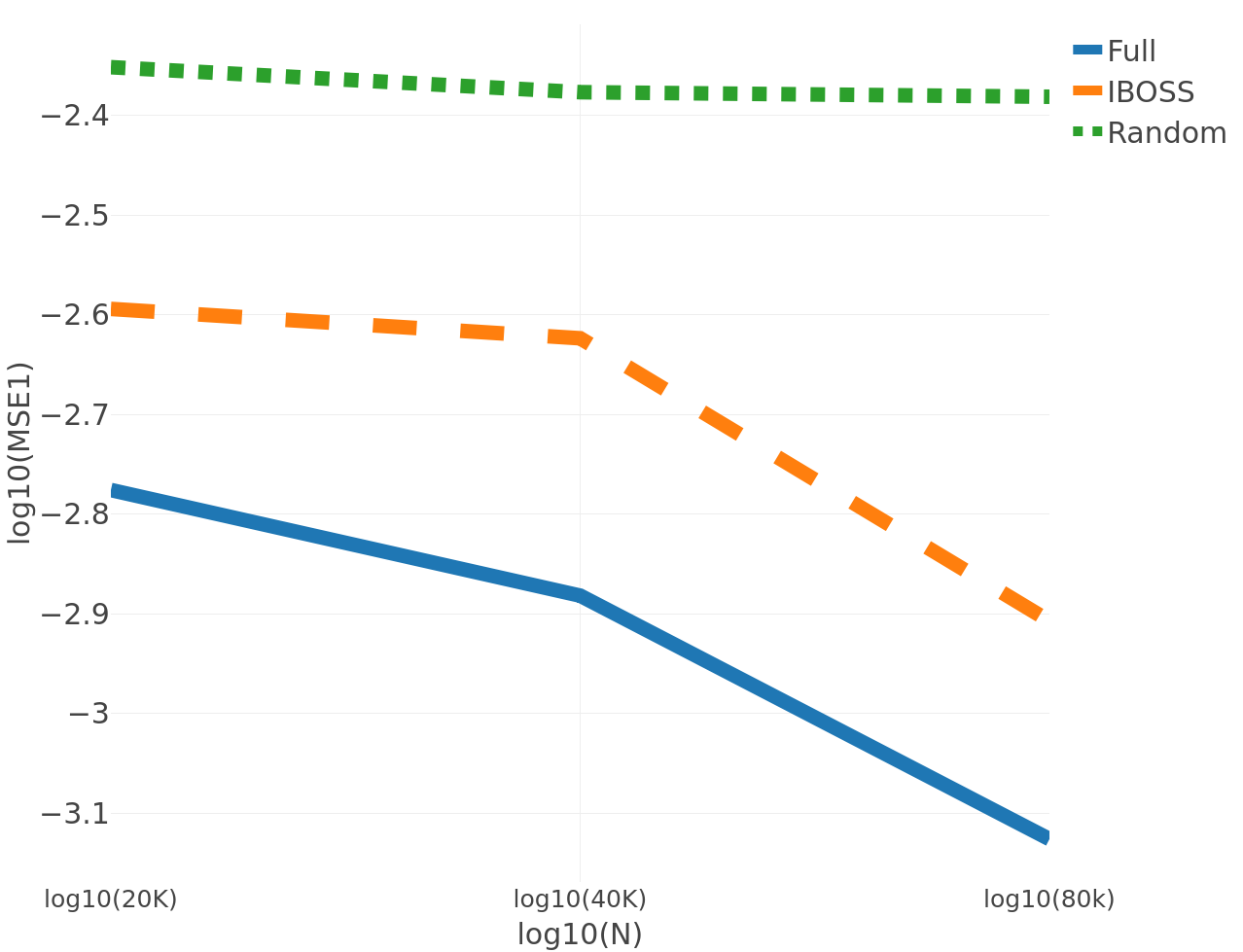}}\hfill
 \subfloat[CPU Time (in seconds)]{\includegraphics[width=.5\linewidth]{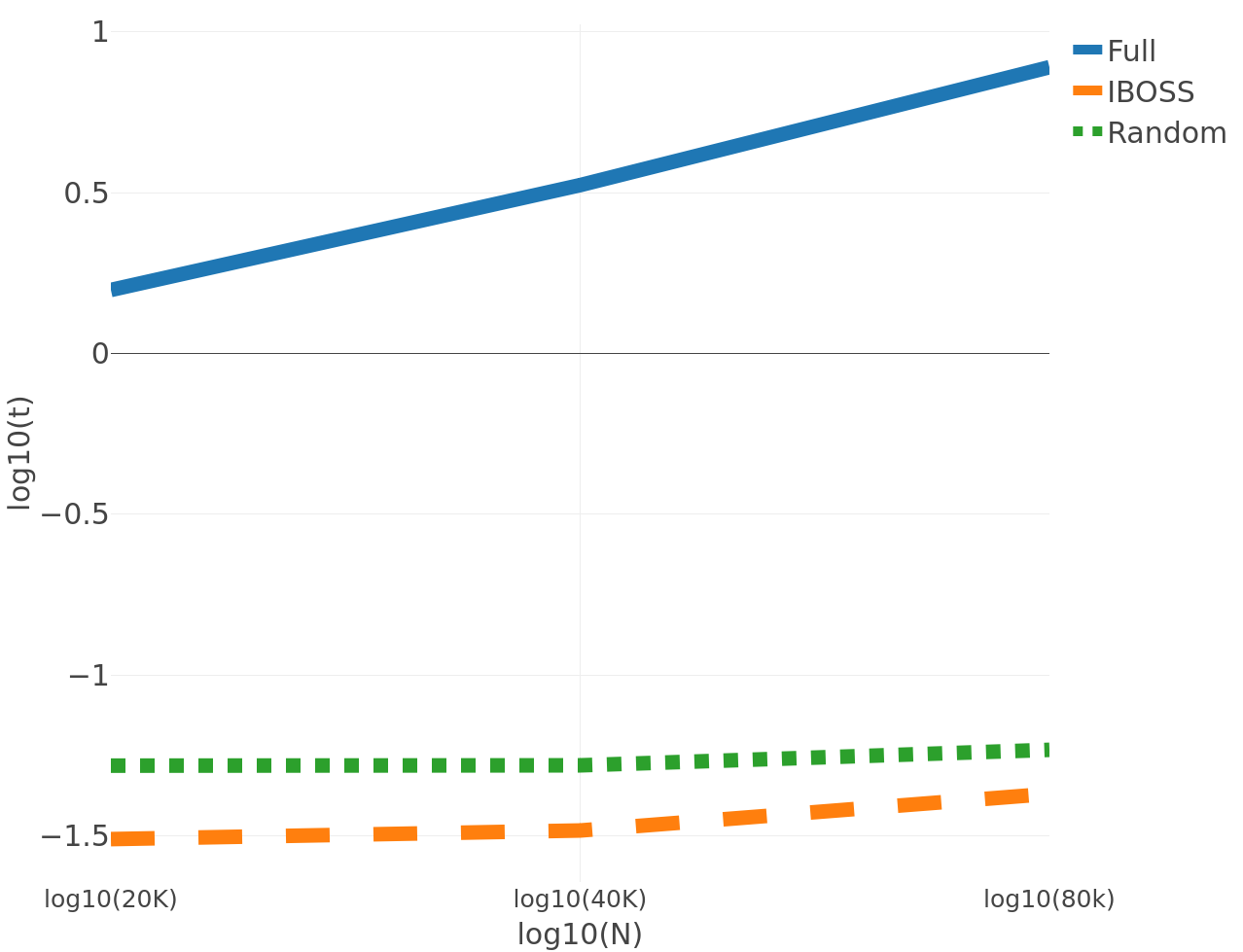}}
 \caption{Comparing different methods for estimating slope parameters based on 500 bootstrap samples of different size $n$ for the Structural Protein Data}
 \label{real_mse1}
 \end{figure}




\begin{figure}
\centering
{\includegraphics[width=0.8\linewidth]{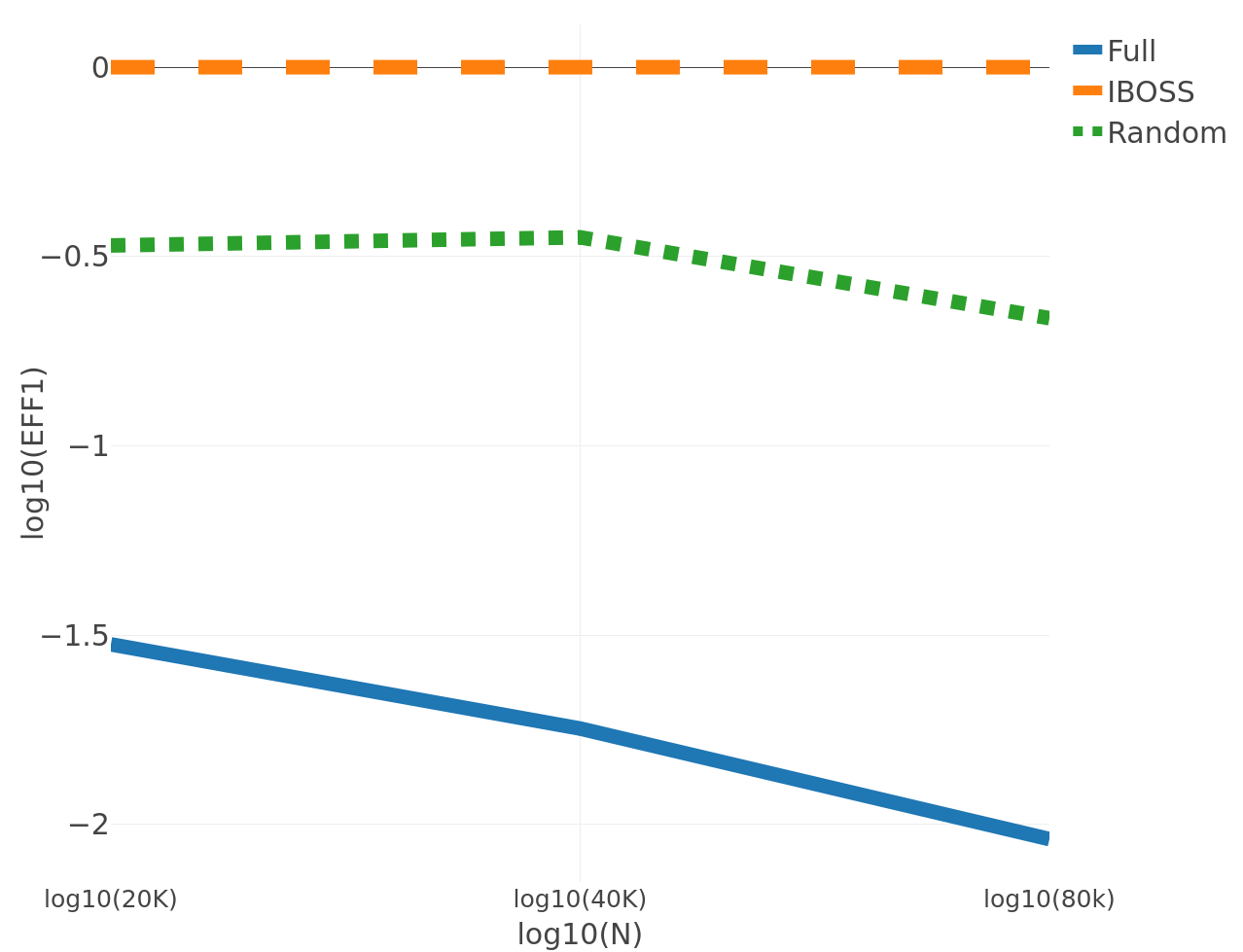}}
\caption{Relative Efficiencies of different methods for slope parameters based on 500 bootstrap samples of different size $n$ for the Structural Protein Data}
\label{real_mse2}
\end{figure}


{Figure~\ref{real_mse1} (a) shows that IBOSS has a smaller MSE for the estimation of slope parameters than Random. Also, as $n$ increases, the MSE for IBOSS decreases, which is consistent with Theorem~\ref{asymp_thm}. For comparing computing time, Figure~\ref{real_mse1} (b) demonstrates a similar pattern as in the simulation studies. Figure~\ref{real_mse2} shows that relative efficiencies for Random and Full tend to decrease when $n$ increases, which is also consistent with results in the simulation studies. }



\section{Conclusions and Future Work}
\label{conclusion}
The size of data sets continues to grow, along with increased heterogeneity in data sets. Mixture-of-Experts (MoE) models are powerful and versatile for modeling and understanding heterogeneous data, but fitting them is computationally expensive, especially for large data sets. One efficient strategy to address this issue is the IBOSS strategy proposed by \citet{Wang2019Information-BasedRegression}. It not only reduces the computational burden by selecting subdata but also retains high statistical efficiency. This paper developed the IBOSS subdata strategy for Clusterwise Linear Regression (CLR) models, a subclass of the MoE models. We proved that, under relatively mild conditions, the IBOSS subdata selection algorithm proposed by \cite{Wang2019Information-BasedRegression} can be used for CLR models. More importantly, we proved that this strategy is asymptotically optimal. The theoretical results are confirmed by simulation studies and a real example. 

There remain important unanswered questions that are beyond the scope of this paper and that need more research. First, different clusters may have different support in the covariate space for a general MoE model with gate functions that depend on the covariates. In this case, IBOSS as applied for CLR models may not work  well. For example, if there is a cluster in which none of the points have any extreme covariate values, we will completely miss that cluster in the subdata. Deriving an IBOSS strategy for general MoE models will be much harder because the more complicated gate functions make the information matrix even more complicated. The path of finding an appropriate matrix that has a closed-form expression and that bounds the actual information matrix could still work, but how to find an appropriate bounding matrix will need additional research. 
Second, the model in each cluster can be a generalized linear regression model or other nonlinear model rather than a linear regression model. This too will make the information matrix and developing an IBOSS subdata selection strategy only more complicated.


While we do not have answers to these questions yet, we expect that these can be resolved in the future by methods akin to those used in this paper. Also, the IBOSS strategy is motivated by results in the optimal design of expriments literature, and we believe that the wealth of knowledge and resources in that literature will continue to provide great guidance for developing innovative and superior subdata techniques and algorithms for general MoE models and many other models.

\emph{Acknowledgments and Funding:}
John Stufken was supported NSF Grant DMS-23-04767 and Min Yang was supported by NSF Grant DMS-22-10546.

\begin{appendix}
\section{The Fisher Information Matrix}
We start with the first derivatives of the log-likelihood with respect to the parameters:
\begin{eqnarray*}
        \frac{\partial{l_{y_i}}}{\partial\pmb{\beta}_g}&=\frac{\pi_g\frac{\partial\phi_{ig}}{\partial\pmb{\beta}_g}}{\sum_{l=1}^G\pi_l\phi_{il}} 
    = w_{ig} \frac{\partial log \phi_{ig}}{\partial \pmb{\beta}_g} \text{ for } g=1,\ldots,G,\\
    \frac{\partial l_{y_i}}{\partial \sigma_g^2}&=\frac{\pi_g\frac{\partial\phi_{ig}}{\partial\sigma_g^2}}{\sum_{l=1}^G\pi_l\phi_{il}} 
    = w_{ig}\frac{\partial log \phi_{ig}}{\partial{\sigma_g^2}} \text{ for } g=1,\ldots,G, \text{ and}\\
    \frac{\partial l_{y_i}}{\partial \pi_g}&=\frac{\phi_{ig}-\phi_{iG}}{\sum_{l=1}^G\pi_l\phi_{il}}=(\frac{w_{ig}}{\pi_g}-\frac{w_{iG}}{\pi_G}) \text{ for } g=1,\ldots,G-1.
\end{eqnarray*}
This leads to the following expressions for the second derivatives of the log-likelihood with respect to the parameters:
\begin{equation*}
\begin{split}
     \frac{\partial^2{l_{y_i}}}{\partial\pmb{\beta}_g \partial \pmb{\beta}_g^T} =& \frac{\partial log \phi_{ig}}{\partial \pmb{\beta}_g}  \frac{\partial w_{ig}}{\partial \pmb{\beta}_g^T} + w_{ig} \frac{\partial^2log\phi_{ig}}{\partial \pmb{\beta}_g\partial \pmb{\beta}_g^T}\\
     =& w_{ig}(1-w_{ig})\frac{\partial log \phi_{ig}}{\partial \pmb{\beta}_g}\frac{\partial log \phi_{ig}}{\partial \pmb{\beta}_g^T}+ w_{ig}  \frac{\partial^2log\phi_{ig}}{\partial \pmb{\beta}_g\partial \pmb{\beta}_g^T},
\end{split}
\end{equation*}
where $\frac{\partial log \phi_{ig}}{\partial \pmb{\beta}_g}=\frac{(y_i-\mb{x_i}^T\pmb{\beta}_g)\mb{x_i}}{\sigma_g^2}$ and $\frac{\partial^2log\phi_{ig}}{\partial \pmb{\beta}_g\partial \pmb{\beta}_g^T}=-\frac{\mb{x_ix_i}^T}{\sigma_g^2}$ for $ 1\leq g  \leq G$,\\
\begin{equation*}
\begin{split}
     \frac{\partial^2{l_{y_i}}}{\partial\sigma_g^2 \partial \sigma_g^2}&=w_{ig}(1-w_{ig})\Bigg[\frac{\partial log \phi_{ig}}{\partial \sigma^2_g}\Bigg]^2+ w_{ig} \cdot \frac{\partial^2log\phi_{ig}}{\partial (\sigma_g^2)^2},
\end{split}
\end{equation*}
where $\frac{\partial log \phi_{ig}}{\partial \sigma_g^2}=-\frac{1}{2\sigma_g^2}+\frac{(y_i-\mb{x_i}^T\pmb{\beta}_g)^2}{2\sigma_g^4}$ and $\frac{\partial^2log\phi_{ig}}{\partial (\sigma_g^2)^2}=\frac{1}{2\sigma_g^4}-\frac{(y_i-\mb{x_i}^T\pmb{\beta}_g)^2}{\sigma_g^6}$ for $ 1\leq g \leq G$, and
\begin{equation*}
\begin{split}
     \frac{\partial^2 l_{y_i}}{(\partial \pi_g)^2}&=-\frac{(\phi_{ig}-\phi_{iG})^2}{(\sum_{l=1}^G\pi_l\phi_{il})^2}, 
\end{split}
\end{equation*}
for $ 1\leq g  \le G-1$.

The Fisher information matrix is now obtained by taking the negative expectation for all second-order derivatives, leading to the form
\begin{equation*}
\mathbi{I}(\mb{x}_i) = 
\begin{pmatrix}
\mathbi{I}_{\pmb{\beta}}(\mb{x}_i)  & \mathbi{I}_{\pmb{\beta,\sigma^2}}(\mb{x}_i) & \mathbi{I}_{\pmb{\beta,\pi}}(\mb{x}_i) \\
\mathbi{I}_{\pmb{\beta,\sigma^2}}^T(\mb{x}_i) & \mathbi{I}_{\pmb{\sigma^2}}(\mb{x}_i) & \mathbi{I}_{\pmb{\sigma^2,\pi}}(\mb{x}_i)\\
\mathbi{I}_{\pmb{\beta,\pi}}^T(\mb{x}_i) &  \mathbi{I}_{\pmb{\sigma^2,\pi}}^T(\mb{x}_i) & \mathbi{I}_{\pmb{\pi}}(\mb{x}_i)
\end{pmatrix}
\end{equation*}
Furthermore,
\begin{equation*}
\mathbi{I}_{\pmb{\beta}}(\mb{x}_i) = 
\begin{pmatrix}
\mathbi{I}_{\pmb{\beta}_1}(\mb{x}_i)  & \mathbi{I}_{\pmb{\beta}_1\pmb{\beta}_2}(\mb{x}_i) & \cdots  &\mathbi{I}_{\pmb{\beta}_1\pmb{\beta}_G}(\mb{x}_i)  \\
\mathbi{I}_{\pmb{\beta}_1\pmb{\beta}_2}(\mb{x}_i) &\mathbi{I}_{\pmb{\beta}_2}(\mb{x}_i) & \cdots & \mathbi{I}_{\pmb{\beta}_2\pmb{\beta}_G}(\mb{x}_i)\\
\vdots & \vdots &\ddots & \vdots\\
 \mathbi{I}_{\pmb{\beta}_1\pmb{\beta}_G}(\mb{x}_i)&\mathbi{I}_{\pmb{\beta}_2\pmb{\beta}_G}(\mb{x}_i)& \cdots & \mathbi{I}_{\pmb{\beta}_G}(\mb{x}_i)
\end{pmatrix}
\end{equation*}
where 
\begin{eqnarray}
\begin{split}
    \mathbi{I}_{\pmb{\beta}_g}(\mb{x}_i)=&-\mathbb{E}\left( w_{ig}(1-w_{ig})\frac{\partial log \phi_{ig}}{\partial \pmb{\beta}_g}\frac{\partial log \phi_{ig}}{\partial \pmb{\beta}_g^T}+ w_{ig}  \frac{\partial^2log\phi_{ig}}{\partial \pmb{\beta}_g\partial \pmb{\beta}_g^T}\right) \\
    =& \pi_g\frac{\mb{x}_i\mb{x}_i^T}{\sigma_g^2}-\mathbb{E}\left( w_{ig}(1-w_{ig})\frac{(y_i-\mb{x_i}^T\pmb{\beta}_g)^2\mb{x_i}\mb{x_i}^T}{\sigma_g^4}\right) 
\end{split}\label{def:ibm}
\end{eqnarray} for $g=1,...,G$;
\begin{equation*}
\mathbi{I}_{\pmb{\sigma^2}}(\mb{x}_i) = 
\begin{pmatrix}
\mathbi{I}_{\sigma^2_1}(\mb{x}_i)  & \mathbi{I}_{\sigma^2_1\sigma^2_2}(\mb{x}_i) & \cdots  &\mathbi{I}_{\sigma^2_1\sigma^2_G}(\mb{x}_i)  \\
\mathbi{I}_{\sigma^2_1\sigma^2_2}(\mb{x}_i) &\mathbi{I}_{\sigma^2_2}(\mb{x}_i) & \cdots & \mathbi{I}_{\sigma^2_2\sigma^2_G}(\mb{x}_i)\\
\vdots & \vdots &\ddots & \vdots\\
 \mathbi{I}_{\sigma^2_1\sigma^2_G}(\mb{x}_i)&\mathbi{I}_{\sigma^2_2\sigma^2_G}(\mb{x}_i)& \cdots & \mathbi{I}_{\sigma^2_G}(\mb{x}_i)
\end{pmatrix}
\end{equation*}
where
\begin{eqnarray}
\begin{split}
\mathbi{I}_{\sigma^2_g}(\mb{x}_i) =& -\mathbb{E}\Bigg\lbrace w_{ig}(1-w_{ig})\Bigg[\frac{\partial log \phi_{ig}}{\partial \sigma^2_g}\Bigg]^2+ w_{ig} \frac{\partial^2log\phi_{ig}}{\partial (\sigma_g^2)^2}\Bigg\rbrace  \\
=& \mathbb{E}\Bigg\lbrace w_{ig}  \Big[ \frac{(y_i-\mb{x_i}^T\pmb{\beta}_g)^2}{\sigma_g^6}-\frac{1}{2\sigma_g^4}\Big]\Bigg\rbrace - \mathbb{E}\Bigg\lbrace w_{ig}(1-w_{ig})\Bigg[\frac{\partial log \phi_{ig}}{\partial \sigma^2_g}\Bigg]^2 \Bigg\rbrace  \\
=& \int_{\mathbb{R}} \frac{\pi_g\phi_{ig}}{\sum_{l=1}^{G}\pi_l\phi_{il}}\Big[ \frac{(y_i-\mb{x_i}^T\pmb{\beta}_g)^2}{\sigma_g^6}-\frac{1}{2\sigma_g^4}\Big] (\sum_{l=1}^{G}\pi_l\phi_{il}) dy_i - \mathbb{E}\Bigg\lbrace w_{ig}(1-w_{ig})\Bigg[\frac{\partial log \phi_{ig}}{\partial \sigma^2_g}\Bigg]^2 \Bigg\rbrace  \\
=& \int_{\mathbb{R}} \pi_g\phi_{ig}\Big[ \frac{(y_i-\mb{x_i}^T\pmb{\beta}_g)^2}{\sigma_g^6}-\frac{1}{2\sigma_g^4}\Big] dy_i - \mathbb{E}\Bigg\lbrace w_{ig}(1-w_{ig})\Bigg[\frac{\partial log \phi_{ig}}{\partial \sigma^2_g}\Bigg]^2 \Bigg\rbrace  \\
=& \frac{\pi_g}{2\sigma_g^4} - \mathbb{E}\Bigg\lbrace w_{ig}(1-w_{ig})\Bigg[-\frac{1}{2\sigma_g^2}+\frac{(y_i-\mb{x_i}^T\pmb{\beta}_g)^2}{2\sigma_g^4}\Bigg]^2\Bigg\rbrace \label{def:ism}
\end{split}
\end{eqnarray}
for $g=1,...,G$; and 
\begin{equation*}
\mathbi{I}_{\pmb{\pi}}(\mb{x}_i) = 
\begin{pmatrix}
\mathbi{I}_{\pi_1}(\mb{x}_i)  & \mathbi{I}_{\pi_1\pi_2}(\mb{x}_i) & \cdots  &\mathbi{I}_{\pi_1\pi_{G-1}}(\mb{x}_i)  \\
\mathbi{I}_{\pi_1\pi_2}(\mb{x}_i) &\mathbi{I}_{\pi_2}(\mb{x}_i) & \cdots & \mathbi{I}_{\pi_2\pi_{G-1}}(\mb{x}_i)\\
\vdots & \vdots &\ddots & \vdots\\
 \mathbi{I}_{\pi_1\pi_{G-1}}(\mb{x}_i)&\mathbi{I}_{\pi_2\pi_{G-1}}(\mb{x}_i)& \cdots & \mathbi{I}_{\pi_{G-1}}(\mb{x}_i)
\end{pmatrix}
\end{equation*}
where
\begin{eqnarray}\label{def:ipm}
\begin{split}
\mathbi{I}_{\pi_g}(\mb{x}_i) =& -\mathbb{E}\Bigg\lbrace-\frac{(\phi_{ig}-\phi_{iG})^2}{(\sum_{g=1}^G\pi_g\phi_{ig})^2}\Bigg\rbrace  \\
=& \mathbb{E}\Bigg\lbrace \frac{w_{ig}^2}{\pi_g^2} + \frac{w_{iG}^2}{\pi_G^2} -2 \frac{w_{ig}w_{iG}}{\pi_g\pi_G} \Bigg\rbrace  \\
=& \frac{1}{\pi_g}+\frac{1}{\pi_G} - \mathbb{E}\Bigg\lbrace \frac{w_{ig}(1-w_{ig})}{\pi_g^2} + \frac{w_{iG}(1-w_{iG})}{\pi_G^2} +2 \frac{w_{ig}w_{iG}}{\pi_g\pi_G} \Bigg\rbrace
\end{split}
\end{eqnarray}for $g=1,...,G-1$.

\section{The proofs of main results}
Before we present a proof of Theorem \ref{surrogate}, we need the following lemma. 
\begin{lemma}
\label{sh_ieq}
Assuming $y_i \sim \sum_{g=1}^G\pi_g\phi(\mb{x}^T_i\mb{\beta}_g,\sigma_g^2)$,  then the following inequalities hold for any $1\leq g_1,g_2\leq G$, $g_1 \not = g_2$:
\begin{equation}
    \begin{split}
&diag\left(\mathbb{E}
\mb{x}_i\mb{x}_i^T{w}_{ig_1}{w}_{ig_2}\Delta_{\pmb{\beta}_{ig_1}}^2
\right) \leq  \frac{1}{2}\mb{f}_{i1}\big(g_1,g_2\big),\\
&\mathbb{E}\left(
{w}_{ig_1}{w}_{ig_2}\Delta_{\pmb{\sigma}_{ig_1}}^2
\right) \leq  \frac{1}{2}f_{i2}\big(g_1,g_2\big),\\
&\mathbb{E}\left(
{w}_{ig_1}{w}_{ig_2}
\right) \leq  \frac{1}{2}{f}_{i3}\big(g_1,g_2\big).
\label{seq:hf_ineq}
    \end{split}
\end{equation} 
Here the first inequality is under the Loewner ordering. 
\end{lemma}
\begin{proof}
Since the proofs of all inequalities are similar, we only provide the proof for the first inequality.
\begin{eqnarray}
&diag\left(\mathbb{E}\left(
\mb{x}_i\mb{x}_i^T{w}_{ig_1}{w}_{ig_2}\Delta_{\pmb{\beta}_{ig_1}}^2
\right)\right)  \nonumber\\
= &diag\left(\mb{x}_i\mb{x}_i^T\int\frac{\pi_{g_1}\phi_{ig_1}\pi_{g_2}\phi_{i{g_2}}}{\sum_{g=1}^G\pi_g\phi_{ig}}\Delta^2_{\pmb{\beta}_{ig_1}} dy_i\right) \nonumber\\
\leq &diag\left(\mb{x}_i\mb{x}_i^T\int\frac{\pi_{g_1}\phi_{ig_1}\pi_{g_2}\phi_{i{g_2}}}{\pi_{g_1}\phi_{ig_1}+\pi_{g_2}\phi_{i{g_2}}}\Delta^2_{\pmb{\beta}_{ig_1}} dy_i\right)\nonumber\\
\leq& diag\left(\mb{x}_i\mb{x}_i^T\int\frac{1}{2}\sqrt{\pi_{g_1}\phi_{ig_1}\pi_{g_2}\phi_{i{g_2}}}\Delta^2_{\pmb{\beta}_{ig_1}} dy_i \right)= \frac{1}{2}\mb{f}_{i1}\big(g_1,g_2\big).\label{seq:57}
\end{eqnarray}
\end{proof}

Now we are ready to prove Theorem \ref{surrogate}.
\begin{proof}[Proof of Theorem \ref{surrogate}]
By (\ref{def:ibm}) and the definition of $\Delta_{\pmb{\beta}_{ig}}$, we have 
\begin{eqnarray*}
diag\left(\mathbi{I}_{\pmb{\beta}_g|M_i}\right)=diag\left(\mb{x_i}\mb{x_i}^T\sum\limits_{g^*:g^* \neq g}\mathbb{E} w_{ig}w_{ig^*} \Delta_{\pmb{\beta}_{ig}}^2\right).
\end{eqnarray*}
Similarly, by (\ref{def:ism}) and the definition of $\Delta_{\pmb{\sigma}_{ig}}$, we have 
\begin{eqnarray*}
\mathbi{I}_{\pmb{\sigma}_g^2|M_i}=\sum\limits_{g^*:g^* \neq g}\mathbb{E} w_{ig}w_{ig^*} \Delta_{\pmb{\sigma}_{ig}}^2
\end{eqnarray*} and 
by (\ref{def:ipm}), we have 
\begin{eqnarray*}
\mathbi{I}_{\pmb{\pi}_g|M_i}=\mathbb{E}\left(\sum\limits_{g^*:g^* \neq g} \left(\frac{w_{ig}w_{ig^*}}{\pi_g^2}+ \frac{w_{iG}w_{ig^*}}{\pi_G^2}\right)+2\frac{w_{ig}w_{iG}}{\pi_g\pi_G}\right).
\end{eqnarray*}
By Lemma \ref{sh_ieq} and the definition of $\pmb{Q}^i$, the conclusion follows. 
\end{proof}

\begin{proof}[Proof of Theorem~\ref{main1}]
By Theorem~\ref{surrogate}, the result follows if we show that $\sum_{i\in \bdelta^*} \mathbi{Q}^i \xrightarrow{\mathbb{P}} \bm{0}_{(Gp+3G-1)\times (Gp+3G-1)}$. This follows if, for all $i \in \bdelta^*$ and $g \not = g'$, 
\begin{eqnarray}
     &\mb{f}_{i1}(g,g')\xrightarrow{\mathbb{P}} \bm{0}_{(p+1)\times (p+1)}, \nonumber\\
    &f_{i2}(g,g')\xrightarrow{\mathbb{P}} 0, \text{ and}\\
    &f_{i3}(g,g')\xrightarrow{\mathbb{P}}, 0 \nonumber\label{t1eq1}
\end{eqnarray}
where $\mb{f}_{i1}$, $f_{i2}$ and $f_{i3}$ are defined in (\ref{eq:f1}) - (\ref{eq:f3}). We prove the two cases separately. 

Case (a):

For any covariate, Algorithm I is guaranteed to select $r$ data points with the $r$ largest values for the covariate in the full data and $r$ data points with the $r$ smallest values of the covariate in the full data. However, when selecting data points based on covariate $l$, $l\ge 2$, some or all of the data points with the $r$ largest and $r$ smallest values for the $l$th covariate may already have been selected. So, Algorithm I may select data points in which none of the values are among the $r$ largest or $r$ smallest values for any covariate. However, what we can guarantee for the subdata $\bdelta^*$ selected by Algorithm I is the following. For any $i \in \bdelta^*$, there exists a $j_i \in \{1,...,p\}$ and $m_i \in \{1,..rp,N-rp+1,...,N\}$ so that $\mb{x}_i=(1, z_{j_i}^{(m_i)1},...,z_{(m_i){j_i}},...,z_{j_i}^{(m_i)p})$, where  $z_{(m_i)j}$ is the $m_i^{th}$ order statistic of $\lbrace z_{1j},...,z_{Nj}\rbrace$ and $z_j^{(m_i)l}$ is the concomitant of $z_{(m_i)j}$ for the $l$th covariate, $l \not = j$. Without loss of generality, let $j_i=1$. For $i=1,...,N$ and $g=1,...,G$, define  $\gamma_{ig}=\mb{x}_i^T\mb{\beta}_g$. Then we have
\begin{eqnarray}
   & \mb{f}_{i1}(g,g')=diag\left( \mb{x}_i\mb{x}_i^T\int\Tilde{w}_i(g,g')\Delta^2_{\bm{\beta}_{ig}} dy_i\right) \nonumber \\
   & = diag\left( \mb{x}_i\mb{x}_i^T\right) \int_\mathbb{R} \sqrt{\pi_g\pi_{g'}}\frac{(y_i-\gamma_{ig})^2}{\sigma_g^4} \frac{1}{\sqrt{2\pi\sigma_g\sigma_{g'}}}\exp\Big\lbrace -\frac{(y_i-\gamma_{ig})^2}{4\sigma_g^2}-\frac{(y_i-\gamma_{i{g'}})^2}{4\sigma_{g'}^2} \Big\rbrace dy_i  \nonumber\\
&= diag\left( \mb{x}_i\mb{x}_i^T\right) \int_\mathbb{R} \sqrt{\pi_g\pi_{g'}}\frac{(y_i-\gamma_{ig})^2}{\sigma_g^4} \frac{1}{\sqrt{2\pi\sigma_g\sigma_{g'}}}\exp\Big\lbrace -\frac{y_i^2-2\frac{\sigma_{g'}^2\gamma_{ig}+\sigma_g^2\gamma_{i{g'}}}{\sigma_g^2+\sigma_{g'}^2}y_i+\frac{\gamma_{ig}^2\sigma_{g'}^2+\gamma_{i{g'}}^2\sigma_g^2}{\sigma_g^2+\sigma_{g'}^2}}{2\cdot \frac{2\sigma_g^2\sigma_{g'}^2}{\sigma_g^2+\sigma_{g'}^2}} \Big\rbrace dy_i \nonumber\\
&= diag\left(\mb{x}_i\mb{x}_i^T\right) \int_\mathbb{R} \sqrt{\pi_g\pi_{g'}}\sqrt{\frac{2\sigma_g\sigma_{g'}}{\sigma_g^2+\sigma_{g'}^2}}\frac{(y_i-\gamma_{ig})^2}{\sigma_g^4} \phi\Big(\frac{\sigma_{g'}^2\gamma_{ig}+\sigma_g^2\gamma_{i{g'}}}{\sigma_g^2+\sigma_{g'}^2},\frac{2\sigma^2_g\sigma^2_{g'}}{\sigma_g^2+\sigma_{g'}^2}\Big)\exp\Big\lbrace -\frac{(\gamma_{ig}-\gamma_{i{g'}})^2}{4(\sigma_g^2+\sigma_{g'}^2)} \Big\rbrace dy_i \nonumber\\
&= diag\left( \mb{x}_i\mb{x}_i^T\right)\sqrt{\frac{2\pi_g\pi_{g'}\sigma_g\sigma_{g'}}{\sigma_g^2+\sigma_{g'}^2}}\Big[ \frac{2\sigma_{g'}^2/\sigma_g^2}{\sigma_g^2+\sigma_{g'}^2}+\frac{(\mb{x}_i^T\bm{\beta}_g-\mb{x}_i^T\bm{\beta}_{g'})^2}{(\sigma_g^2+\sigma_{g'}^2)^2} \Big]\exp\Big\lbrace -\frac{(\mb{x}_i^T\bm{\beta}_g-\mb{x}_i^T\bm{\beta}_{g'})^2}{4(\sigma_g^2+\sigma_{g'}^2)}\Big\rbrace \nonumber\\
&=diag\left( \mb{x}_i\mb{x}_i^T\right)\sqrt{\frac{2\pi_g\pi_{g'}\sigma_g\sigma_{g'}}{\sigma_g^2+\sigma_{g'}^2}}\Big[ \frac{2\sigma_{g'}^2/\sigma_g^2}{\sigma_g^2+\sigma_{g'}^2}+\frac{\big(\beta_{g,0}-\beta_{{g'},0}+ z_{(m_i)1}(\beta_{g,1}-\beta_{{g'},1})+\sum_{l=2 }^pz_{1}^{(m_i)l}(\beta_{g,l}-\beta_{{g'},l})\big)^2}{(\sigma_g^2+\sigma_{g'}^2)^2} \Big]\times \nonumber\\
&\exp\Big\lbrace -\frac{\big(\beta_{g,0}-\beta_{{g'},0}+ z_{(m_i)1}(\beta_{g,1}-\beta_{{g'},1})+\sum_{l=2}^pz_{1}^{(m_i)l}(\beta_{g,l}-\beta_{{g'},l})\big)^2}{4(\sigma_g^2+\sigma_{g'}^2)}\Big\rbrace
\label{eq:10}
\end{eqnarray}
where $\mb{x}_i^T=(1,z_{(m_i){1}},z_{1}^{(m_i)2}...,z_{1}^{(m_i)p})$. From the results in Examples 2.8.1 and 5.5.1 of \cite{Galambos1987TheStatistics},
when $(z_{i1},...,z_{ip}) \sim \mathbi{N}(\bm{\mu}_z,\bm{\Sigma}_z)$
\begin{eqnarray}
 z_{(m_i)1}=\mu_{z1}-\sigma_{z1}\sqrt{2logN}+O_P(1),\quad &m_i=1,...rp, \label{eq:11}\\
 z_{(m_i)1}=\mu_{z1}+\sigma_{z1}\sqrt{2logN}+O_P(1),\quad &m_i=N-rp+1,...,N, \label{eq:12}\\
  z_{1}^{(m_i)l}=\mu_{z1}-\rho_{l1}\sigma_{z1}\sqrt{2logN}+O_P(1),\quad &m_i=1,...rp,\label{eq:13}\\
 z_{1}^{(m_i)l}=\mu_{z1}+\rho_{l1}\sigma_{z1}\sqrt{2logN}+O_P(1),\quad &m_i=N-rp+1,...,N.\label{eq:14}
\end{eqnarray}
We distinguish between $m_i \in \{1,...,rp\}$ and $m_i \in \{N-rp+1,...,N\}$. First, for $m_i \in \{1,...rp\}$, by (\ref{eq:11}) and (\ref{eq:13}) we have $\mb{x}_i^T=\bigg(1, -\rho_{11}\sigma_{z1}\sqrt{2logN}+O_p(1),\dots,-\rho_{p1}\sigma_{z1}\sqrt{2logN}+O_p(1)\bigg)$, so that (\ref{eq:10}) can be written as
\begin{eqnarray}
& diag\left( \mb{x}_i\mb{x}_i^T\right)\sqrt{\frac{2\pi_g\pi_{g'}\sigma_g\sigma_{g'}}{\sigma_g^2+\sigma_{g'}^2}}\Big[ \frac{2\sigma_{g'}^2/\sigma_g^2}{\sigma_g^2+\sigma_{g'}^2}+\frac{\big(-\sqrt{2logN}\sum\limits_{l=1}^p\rho_{l1}\sigma_{z1}(\beta_{g,l}-\beta_{g',l})+O_p(1)\big)^2}{(\sigma_g^2+\sigma_{g'}^2)^2} \Big]\times \nonumber\\
&\exp\Big\lbrace -\frac{\big(-\sqrt{2logN}\sum\limits_{l=1}^p\rho_{l1}\sigma_{z1}(\beta_{g,l}-\beta_{g',l})+O_p(1)\big)^2}{4(\sigma_g^2+\sigma_{g'}^2)}\Big\rbrace. \label{eq:15}
\end{eqnarray}
Second, for $m_i \in \{N-rp+1,...,N\}$, by (\ref{eq:12}) and (\ref{eq:14}) we have $\mb{x}_i=\bigg(1, \rho_{11}\sigma_{z1}\sqrt{2logN}+O_p(1),\dots,\rho_{p1}\sigma_{z1}\sqrt{2logN}+O_p(1)\bigg)$, so that (\ref{eq:10}) can be written as
\begin{eqnarray}
& diag\left( \mb{x}_i\mb{x}_i^T\right)\sqrt{\frac{2\pi_g\pi_{g'}\sigma_g\sigma_{g'}}{\sigma_g^2+\sigma_{g'}^2}}\Big[ \frac{2\sigma_{g'}^2/\sigma_g^2}{\sigma_g^2+\sigma_{g'}^2}+\frac{\big(\sqrt{2logN}\sum\limits_{l=1}^p\rho_{l1}\sigma_{z1}(\beta_{g,l}-\beta_{g',l})+O_p(1)\big)^2}{(\sigma_g^2+\sigma_{g'}^2)^2} \Big]\times \\
&\exp\Big\lbrace -\frac{\big(\sqrt{2logN}\sum\limits_{l=1}^p\rho_{l1}\sigma_{z1}(\beta_{g,l}-\beta_{g',l})+O_p(1)\big)^2}{4(\sigma_g^2+\sigma_{g'}^2)}\Big\rbrace. \label{eq:16}
\end{eqnarray}
With the condition for Case (a), $\sum\limits_{l=1}^p\rho_{l1}\sigma_{z1}(\beta_{g,l}-\beta_{g',l}) \neq 0$, this implies that when $N \rightarrow \infty$, $\eqref{eq:15} \xrightarrow{\mathbb{P}} \bm{0}_{(p+1)\times (p+1)}$ and $\eqref{eq:16} \xrightarrow{\mathbb{P}} \bm{0}_{(p+1)\times (p+1)}$. Consequently $\mb{f}_{i1}(g,g')\xrightarrow{\mathbb{P}} \bm{0}_{(p+1)\times (p+1)}$.
\\

Case (b):
By the same argument as in the proof of Case (a), it suffices to show that, for all $ i \in \bdelta^*$, 
\begin{eqnarray}
\begin{split}
    & \mb{f}_{i1}\big(g,g')\xrightarrow{\mathbb{P}} \pmb{0}_{(p+1)\times (p+1)} \\
    &f_{i2}(g,g')\xrightarrow{\mathbb{P}} 0\\
    &f_{i3}(g,g')\xrightarrow{\mathbb{P}} 0\label{t2eq1}
\end{split}
\end{eqnarray}
for any pair $(g,g')$. Since proofs of the three convergences are similar, we only show a proof of the first one and use the same notation as in the proof for part (a) of Theorem \ref{main1}. Without loss of generality, set $j_i=1$. By the same argument as used in (\ref{eq:10}), we have
 \begin{eqnarray}
  \mb{f}_{i1}(g,g')=&diag\left( \mb{x}_i\mb{x}_i^T\right)\sqrt{\frac{2\pi_g\pi_{g'}\sigma_g\sigma_{g'}}{\sigma_g^2+\sigma_{g'}^2}}\Big[ \frac{2\sigma_{g'}^2/\sigma_g^2}{\sigma_g^2+\sigma_{g'}^2}+\frac{\big(\beta_{g,0}-\beta_{{g'},0}+ z_{(m_i)1}(\beta_{g,1}-\beta_{{g'},1})+\sum_{l=2 }^pz_{1}^{(m_i)l}(\beta_{g,l}-\beta_{{g'},l})\big)^2}{(\sigma_g^2+\sigma_{g'}^2)^2} \Big]\times \nonumber\\
&\exp\Big\lbrace -\frac{\big(\beta_{g,0}-\beta_{{g'},0}+ z_{(m_i)1}(\beta_{g,1}-\beta_{{g'},1})+\sum_{l=2}^pz_{1}^{(m_i)l}(\beta_{g,l}-\beta_{{g'},l})\big)^2}{4(\sigma_g^2+\sigma_{g'}^2)}\Big\rbrace,
\label{eq:17}
 \end{eqnarray} where $\mb{x}_i^T=(1,z_{(m_i){1}},z_{1}^{(m_i)2}...,z_{1}^{(m_i)p})$.
 From the results in Theorem 6 of \cite{Wang2019Information-BasedRegression}, when $(z_{i1},...,z_{ip}) \sim LN(\pmb{\mu}_z,\pmb{\Sigma}_z)$, 
\begin{eqnarray}
&z_{(m_i)1}=exp\big(-\sigma_{z1}\sqrt{2log N}\big)O_P(1),\quad &m_i \in \{1,...,rp\}; \label{eq:18}\\
&z_{(m_i)1}=exp\big(\sigma_{z1}\sqrt{2log N}\big)O_P(1),\quad &m_i \in \{N-rp+1,...,N\}; \label{eq:19}\\
&z_{1}^{(m_i)l}=exp\big(-\rho_{l1}\sigma_{z1}\sqrt{2log N}\big)O_P(1),\quad &m_i \in \{1,...,rp\};\label{eq:20}\\
&z_{1}^{(m_i)l}=exp\big(\rho_{l1}\sigma_{z1}\sqrt{2log N}\big)O_P(1),\quad &m_i \in \{N-rp+1,...,N\}.\label{eq:21}
\end{eqnarray}
As in the proof for Case (a), we consider the cases $m_i \in \{1,...,rp\}$ and $m_i \in \{N-rp+1,...,N\}$. First, for $m_i \in \{1,...,rp\}$, by (\ref{eq:18}) and (\ref{eq:20}), (\ref{eq:17}) can be written as
\begin{eqnarray}
&diag\left(\mb{x}_i\mb{x}_i^T\right)\sqrt{\frac{2\pi_g\pi_{g'}\sigma_g\sigma_{g'}}{\sigma_g^2+\sigma_{g'}^2}}\Big[ \frac{2\sigma_{g'}^2/\sigma_g^2}{\sigma_g^2+\sigma_{g'}^2}+ \frac{A_1^2}{(\sigma_g^2+\sigma_{g'}^2)^2} \Big]\times \exp\Big\lbrace -\frac{A_1^2}{4(\sigma_g^2+\sigma_{g'}^2)}\Big\rbrace,
\label{eq:22}
\end{eqnarray}
where 
\begin{eqnarray*}
\mb{x}_i=&\bigg(1,\quad exp\big\lbrace-\rho_{11}\sigma_{z1}\sqrt{2log N}\big\rbrace O_P(1)\quad,\dots,\quad exp\big\lbrace-\rho_{p1}\sigma_{z1}\sqrt{2log N}\big\rbrace O_P(1)\bigg) \text{ and}\\
A_1=&\beta_{g,0}-\beta_{g',0}+O_P(1)\Bigg[exp\Big\lbrace-\rho_{\min,1}\sigma_{z1}\sqrt{2log N} \Big\rbrace \times \sum\limits_{l \in \mathcal{L}_{\min,1}}\big(\beta_{g,l}-\beta_{g',l}\big)+  \nonumber\\
&\sum\limits_{l \notin \mathcal{L}_{\min,1}}\Big( exp\Big\lbrace -\rho_{lj}\sigma_{z1}\sqrt{2log N} \Big\rbrace(\beta_{g,l}-\beta_{g',l})\Big)\Bigg]. 
\end{eqnarray*}
With the condition on the parameters for Case (b), we have that $\rho_{\min,j} < 0$ and $ \sum\limits_{l \in \mathcal{L}_{\min,j}}\big(\beta_{g,l}-\beta_{g',l}\big) \neq 0$. Thus  $\eqref{eq:22} \xrightarrow{\mathbb{P}} \pmb{0}_{(p+1)\times (p+1)}$ when $N \rightarrow \infty$.\\
Second, for $m_i \in \{N-rp+1, ..., N\}$, by \eqref{eq:19} and \eqref{eq:21}, \eqref{eq:17} can be written as
\begin{eqnarray}
&diag\left( \mb{x}_i\mb{x}_i^T\right)\sqrt{\frac{2\pi_g\pi_{g'}\sigma_g\sigma_{g'}}{\sigma_g^2+\sigma_{g'}^2}}\Big[ \frac{2\sigma_{g'}^2/\sigma_g^2}{\sigma_g^2+\sigma_{g'}^2}+ \frac{A_2^2}{(\sigma_g^2+\sigma_{g'}^2)^2} \Big]\times \exp\Big\lbrace -\frac{A_2^2}{4(\sigma_g^2+\sigma_{g'}^2)}\Big\rbrace
\label{eq:25}
\end{eqnarray}
where 
\begin{eqnarray*}
\mb{x}_i=&\bigg(1,\quad exp\big\lbrace\rho_{11}\sigma_{z1}\sqrt{2log N}\big\rbrace O_P(1)\quad,\dots,\quad exp\big\lbrace\rho_{p1}\sigma_{z1}\sqrt{2log N}\big\rbrace O_P(1)\bigg) \text{ and}\\
A_2=&\beta_{g,0}-\beta_{g',0}+O_P(1)\Bigg[exp\Big\lbrace\sigma_{z1}\sqrt{2log N} \Big\rbrace \times \big(\beta_{g,1}-\beta_{g',1}\big)+  \nonumber\\
&\sum\limits_{l>1}\Big( exp\Big\lbrace \rho_{l1}\sigma_{z1}\sqrt{2log N} \Big\rbrace(\beta_{g,l}-\beta_{g',l})\Big)\Bigg] 
\end{eqnarray*}
With the condition on the parameters for Case (b), we have  $ \beta_{g,1}-\beta_{g',1} \neq 0$. Thus $\eqref{eq:25} \xrightarrow{\mathbb{P}} \pmb{0}_{(p+1)\times (p+1)}$ when $N \rightarrow \infty$. Thus the conclusion follows.

\end{proof}

\begin{proof}[Proof of Theorem \ref{asymp_thm}]

For Case (a), by Theorem 6 in \cite{Wang2019Information-BasedRegression}, when $\mb{z}_i \sim
N(\bm{\mu}_z, \bm{\Sigma}_z)$,
\begin{eqnarray}
\sum_{i\in \bdelta^*} \mb{x}_i\mb{x}_i^T=
 \begin{pmatrix}
 k & \bm{0}\\
 \bm{0} & ~~4r\log N\bm{\Phi}_z\bm{\rho}^2\bm{\Phi}_z
\end{pmatrix}+O_P(\sqrt{\log N}) \label{prf_thm4_a1}
\end{eqnarray}
and
\begin{eqnarray}
\bm{A}_N\left(\sum_{i\in \bdelta^*} \mb{x}_i\mb{x}_i^T\right)^{-1}\bm{A}_N=
 \begin{pmatrix}
 \frac{1}{k} &  \bm{0}\\
 \bm{0} & ~~\frac{1}{4r}(\bm{\Phi}_z\bm{\rho}^2\bm{\Phi}_z)^{-1}
\end{pmatrix}+O_P\Big(\frac{1}{(\sqrt{\log N}}\Big).\label{prf_thm4_a2}
\end{eqnarray}
Notice that $\mathbi{I}(\bdelta^*)=\sum_{i\in \bdelta^*}\mathbi{I}_{C_i} - \sum_{i\in \bdelta^*} \mathbi{I}_{M_i}$. 
By Theorems \ref{surrogate} and \ref{main1}, we have 
 $\sum_{i\in \bdelta^*} \mathbi{I}_{M_i} \xrightarrow{\mathbb{P}} \bm{0}_{(Gp+3G-1)\times (Gp+3G-1)}$ when $N\rightarrow \infty$, which implies that
 $\mathbi{I}(\bdelta^*)\xrightarrow{\mathbb{P}} \sum_{i\in \bdelta^*}\mathbi{I}_{C_i}
$  when $N\rightarrow \infty$. 
By the expressions for  
$\mathbi{I}_{C_i}$ and  $\mathbi{I}_{\bm{\beta}|C_i}$ in \eqref{complete_infor} and \eqref{infor_beta_c}, respectively, the desired conclusion follows from (\ref{prf_thm4_a2}).\\

For Case (b), also by Theorem 6 in \cite{Wang2019Information-BasedRegression}, when $\mb{z}_i \sim LN(\bm{\mu}_z,\bm{\Sigma}_z)$,
\begin{eqnarray}
\sum_{i\in \bdelta^*} \mb{x}_i\mb{x}_i^T=
\begin{pmatrix}
 k & \mb{v}^T\\
 \mb{v} & \bm{\Omega},
\end{pmatrix} \label{prf_thm4_b1}
\end{eqnarray}
where, with $\mb{v}^T=(v_1,\dots,v_p)$ and $\bm{\Omega}=\left( \Omega_{j_1j_2} \right)_{p \times p}$,
\begin{eqnarray*}
\Omega_{jj}= r\exp\Big(2\sigma_{zj}\sqrt{2\log N}\Big)\Big\lbrace e^{2\mu_{zj}}+o_p(1)\Big\rbrace,\\
\Omega_{j_1j_2}= 2r\exp\Big\lbrace(\sigma_{zj_1}+\sigma_{zj_2})\sqrt{2\log N}\Big\rbrace o_p(1), \text{ and}\\
v_j=r\exp\Big(\sigma_{zj}\sqrt{2\log N}\Big)\Big\lbrace e^{\mu_{zj}}+o_p(1)\Big\rbrace
\end{eqnarray*}
and
\begin{eqnarray}
\bm{B}_N\left({\sum_{i\in \bdelta^*} \mb{x}_i\mb{x}_i^T}\right)^{-1}\bm{B}_N=\frac{2}{k}\begin{pmatrix}
 1 & -\bm{\nu}^T\\
-\bm{\nu} & ~~p\bm{\Psi}+\bm{\nu}\bm{\nu}^T 
\end{pmatrix}+o_P(1). \label{prf_thm4_b2}
\end{eqnarray}
By a similar argument as for Case (a), the desired conclusion follows.

Next we want to show that $\bdelta^*$ provides the fastest convergence rate for $V(\hat{\beta}_{g,j}^{\bdelta})\xrightarrow{\mathbb{P}} 0$ among all subdata $\bdelta$ of size $k$. We consider Case (a) only since the proof for Case (b) is similar. 
From (\ref{eq:2}), for any $\bdelta$ with subdata size $k$, we have $\mathbi{I}({\bdelta})^{-1} \geq \left(\sum_{i\in \bdelta}\mathbi{I}_{C_i}\right)^{-1}$ in Loewner order, and further we have $\lbrace\mathbi{I}({\bdelta})^{-1}\rbrace_{jj} \geq \lbrace\left(\sum_{i\in \bdelta}\mathbi{I}_{C_i}\right)^{-1}\rbrace_{jj}\geq \left(\lbrace \sum_{i\in \bdelta}\mathbi{I}_{C_i}\rbrace_{jj}\right)^{-1}$ for all j. Then for estimating the slope parameters of the $g${th} cluster with any subdata $\bdelta$, we have
\begin{equation}
    \begin{split}
V(\hat{\beta}_{g,j}^{\bdelta}) &\geq \frac{\sigma_g^2}{\pi_g}(\sum\limits_{i \in \bdelta}z_{ij}^2)^{-1} \geq \frac{\sigma_g^2}{\pi_g}\min\left((kz_{(1)j}^2)^{-1},(kz_{(N)j}^2)^{-1}\right)\\
&=\frac{\sigma_g^2}{k\pi_g}\min\left((\mu_{z1}+\sigma_{z1}\sqrt{2logN}+o_P(1))^{-2},(\mu_{z1}+\sigma_{z1}\sqrt{2logN}+o_P(1))^{-2}\right)   \label{asym_rate_1}     
    \end{split}
\end{equation}
for $j=1,...,p$. From \eqref{asym_rate_1}, for any $\bdelta$, the lower bound of the convergence rate of  $V(\hat{\beta}_{g,j}^{\bdelta})$ is $1/\log N$. On the other hand, from \eqref{asymp_1}, it is clear $V(\hat{\beta}_{g,j}^{\bdelta^*})$ achieves  this lower bound. 
\end{proof}

\end{appendix}

\bibliographystyle{natbib}
\bibliography{references}     
\end{document}